\documentclass[a4paper,twocolumn,10pt,notitlepage]{article}
\usepackage[utf8]{inputenc}
\usepackage[pdftex]{graphicx}
\usepackage[papersize={223mm,297mm},t   extheight=260mm,textwidth=190mm]{geometry}
\usepackage{float} 
\usepackage{amsmath}
\usepackage{braket}
\usepackage{multirow}
\usepackage{array}
\usepackage{xcolor}
\usepackage{caption}
\usepackage[Symbol]{upgreek}
\usepackage[affil-it]{authblk}
\usepackage{siunitx}
\usepackage[
backend=biber,
style=nature,
doi=false,
isbn=false,
url=false,
eprint=false
]{biblatex}
\DeclareCaptionFormat{custom}
{%
    \textbf{#1#2}#3
}
\usepackage{xr}
\makeatletter
\newcommand*{\addFileDependency}[1]{
  \typeout{(#1)}
  \@addtofilelist{#1}
  \IfFileExists{#1}{}{\typeout{No file #1.}}
}
\makeatother

\newcommand*{\myexternaldocument}[1]{%
    \externaldocument{#1}%
    \addFileDependency{#1.tex}%
    \addFileDependency{#1.aux}%
}
\myexternaldocument{Supplement}

\title{ARPES signatures of few-layer twistronic graphenes}

\author[1,2]{J. E. Nunn*}
\author[3,4]{A. McEllistrim*}
\author[3,4]{A. Weston*}
\author[3,4]{A. Garcia-Ruiz}
\author[1]{M. D. Watson}
\author[5]{M. Mucha-Kruczynski}
\author[1]{C. Cacho}
\author[3,4]{R. Gorbachev}
\author[3,4]{V.I. Fal'ko}
\author[2]{N.R. Wilson}

\affil[1]{Diamond Light Source, Division of Science, Didcot, OX11 0DE, UK}
\affil[2]{Department of Physics, University of Warwick, Coventry CV4 7AL, UK}
\affil[3]{School of Physics and Astronomy, University of Manchester, Oxford Road, Manchester M13 9PL, UK}
\affil[4]{National Graphene Institute, University of Manchester, Booth St East, Manchester M13 9PL, UK}
\affil[5]{Centre for Nanoscience and Nanotechnology, Department of Physics, University of Bath, Bath, BA2 7AY, UK}

\date{* These authors contributed equally to this work. E-mail: Vladimir.Falko@Manchester.ac.uk, Roman@Manchester.ac.uk, Cephise.Cacho@Diamond.ac.uk and Neil.Wilson@Warwick.ac.uk}

\begin{document}

\twocolumn[{
\maketitle
\begin{@twocolumnfalse}
\begin{abstract}
\noindent

Diverse emergent correlated electron phenomena have been observed in twisted graphene layers due to electronic interactions with the moiré superlattice potential. Many electronic structure predictions have been reported exploring this new field, but with few momentum-resolved electronic structure measurements to test them. Here we use angle-resolved photoemission spectroscopy (ARPES) to study the twist-dependent ($1^\circ < \theta < 8^\circ$) electronic band structure of few-layer graphenes, including twisted bilayer, monolayer-on-bilayer, and double-bilayer graphene (tDBG). Direct comparison is made between experiment and theory, using a hybrid $\textbf{k}\cdot\textbf{p}$ model for interlayer coupling and implementing photon-energy-dependent phase shifts for photo-electrons from consecutive layers to simulate ARPES spectra. Quantitative agreement between experiment and theory is found across twist angles, stacking geometries, and back-gate voltages, validating the models and revealing displacement field induced gap openings in twisted graphenes. However, for tDBG at $\theta=1.5\pm0.2^\circ$, close to the predicted magic-angle of $\theta=1.3^\circ$, a flat band is found near the Fermi-level with measured bandwidth of $E_w = 31\pm5$ meV. Analysis of the gap between the flat band and the next valence band shows significant deviations between experiment ($\Delta_h=46\pm5$~meV) and the theoretical model ($\Delta_h=5$~meV), indicative of the importance of lattice relaxation in this regime.

\end{abstract}
\vspace{3mm}
\end{@twocolumnfalse}
}]

\textbf{Introduction}

Reports of anomalous superconductivity~\cite{Cao2018SC} and correlated insulating behaviour~\cite{Cao2018Mott} in magic angle twisted bilayer graphene (MATBG) sparked an avalanche of research into magic-angle effects in two-dimensional materials (2DMs) and into 2D twistronics more generally. Overlapping two identical 2D crystal lattices with a small angular rotation (twist angle, $\theta$) between them creates a long-range moiré superlattice. For MATBG, moiré interactions between the layers create a flat band near the Fermi level~\cite{Bistritzer2011} whose filling can be electrostatically controlled by gate electrodes. The high density of states within this flat band results in strong, and gate-tunable, electron correlation effects~\cite{Cao2018SC,Cao2018Mott,Choi2019} as also observed in twisted bilayer transition metal dichalcogenides~\cite{Wang2020,Zhang2020} and in twisted few-layer graphenes~\cite{Chen2021,Xu2021,Cao2020,Shen2020,Zhang2021,Liu2021}. However, there are challenges to modelling these systems. The large number of atoms in a moiré cell complicate the application of \textit{ab initio} approaches, leading to the development of various multiscale approaches~\cite{Carr2020} such as large-scale density functional theory~\cite{dft1, dft2, Lucignano2019}, tight-binding and continuum models~\cite{dosSantos2007,Bistritzer2011, GarciaRuiz2021}. Although these give qualitatively similar predictions, the details of the dispersions, and hence their properties, depend on the simulation methodology and parameter set used. Experimental studies are therefore vital to validate and refine the theoretical models and to understand the electronic band structure changes which underlie the emergent twistronic effects.

Angle-resolved photoemission spectroscopy (ARPES) gives unique insight into the momentum-resolved electronic band structure of 2DMs and 2D heterostructures~\cite{Bostwick2007,Riley2015,Kim2015,Tang2017,Wilson2017,Katoch2018,Jin2015}. Due to the short mean-free path of the photo-excited electrons, ARPES is sensitive to the top few atomic layers, enabling the study of layer-dependent effects, while \textit{in-situ} back gating of 2D heterostructures during ARPES allows the study of band structure changes with carrier concentration~\cite{Nguyen2019,Joucken2019,Muzzio2020} and with transverse displacement fields~\cite{Jones2020,Nguyen2021}. ARPES has previously been applied to the study of twisted graphenes, initially studying multi-layer graphene grown on SiC or copper where twisted regions can be found by chance~\cite{Kandyba2015,Razado2016,Peng2017,Thompson_2020,Iimori2021}. However, interactions with the substrate cause complications such as inhomogeneous doping, increased screening, and additional moiré periodicities. Instead, mechanical exfoliation and stacking on boron nitride can be used to fabricate twisted graphene samples at defined twist angles for ARPES, for example, showing that moiré superlattice effects persist even in large-angle twisted bilayer graphene where the moiré period is short~\cite{Hamer2022}. Also using this approach, recent reports of ARPES on MATBG detected a flat band at the Fermi-level~\cite{Utama2021,Lisi2021}, although neither the flat band dispersion nor the Fermi surface topology could be resolved. 

\begin{figure*}[t]
    \centering
    \includegraphics[width=0.785\textwidth]{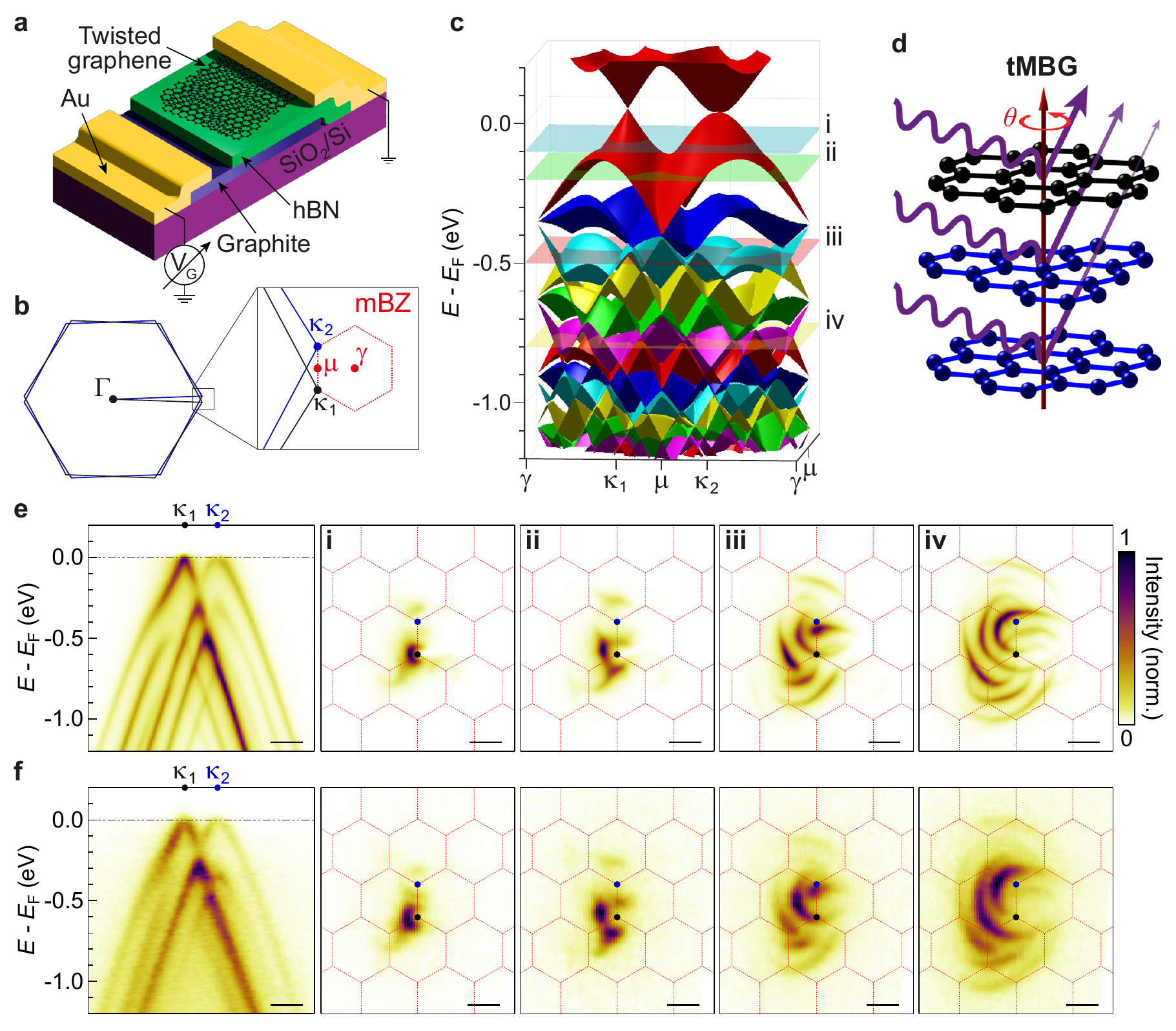}
    \caption{\textbf{Comparison of simulated and measured ARPES spectra for tMBG}. Schematic of \textbf{a}~the twisted graphene heterostructures and \textbf{b}~the twisted graphene Brillouin zones and corresponding moiré Brillouin zone (mBZ). \textbf{c}~Band structure for tMBG at $\theta =$~3.4$^\circ$ computed using the hybrid $\textbf{k} \cdot \textbf{p}$ model. \textbf{d}~Schematic of the photoemission process in multilayered graphene. Photoelectrons from deeper layers are attenuated due to scattering in the material. \textbf{e}~Simulated and \textbf{f}~experimental ARPES spectra for tMBG at $\theta =$~3.4$^\circ$. Left-hand panels show energy-momentum cuts taken along the $\upkappa_{1} \rightarrow \upkappa_{2}$ direction shown in \textbf{b}, where $\upkappa_{1}$ corresponds to the upper layer. Panels \textbf{i-iv} are constant energy maps at $\textit{E}-\textit{E}_\text{F}=$ \textbf{i}~-100~meV, \textbf{ii}~-200~meV, \textbf{iii}~-500~meV and \textbf{iv}~-800~meV, as marked by the horizontal planes in \textbf{d}. All scale bars are 0.1~\AA$^{-1}$.}
    \label{fig1}
\end{figure*}

Here, we use direct comparison between simulated and measured ARPES spectra to test electronic band structure predictions for few-layer graphene samples with different stacking geometries and over a range of (small) twist angles. Our measurements visualise the twist-dependent electronic band structure, giving a quantitative test of the validity of the hybrid $\textbf{k} \cdot \textbf{p}$ model and the corresponding choice of empirical parameters. We measure the dispersion of the flat valence band near the Fermi-level in twisted double bilayer graphene, finding small but significant differences between the experimental results and the hybrid $\textbf{k}\cdot\textbf{p}$ model simulations. Extending this, using ARPES with \textit{in-situ} gating we show that, away from this magic angle regime, the model accurately describes the gate-dependent behaviour: applying a back-gate voltage results in both electrostatic doping of the graphene layers and a displacement field across them, that opens a field-dependent gap at the Dirac point of bilayer graphene.

\textbf{Results and discussion}

Twisted few layer graphene samples were fabricated on hexagonal boron nitride (hBN) by a modified tear-and-stack approach~\cite{Frisenda2018, Kim2016},  as described in detail in Supplementary Information (SI) section~1 and Methods. A graphite back gate electrode was incorporated into some of the devices, shown schematically in Fig.~\ref{fig1}a. Scanning probe microscopy and scanning photoemission microscopy showed homogeneous regions a few micrometres across in the twisted graphenes. Spatially resolved ARPES spectra were acquired from within these regions at the nanoARPES branch of the I05 beamline at Diamond Light Source, as described in Methods and SI section~2. Energies are measured relative to the Fermi energy, $E_\text{F}$, determined by fitting the drop in photoemission intensity at $E_\text{F}$ on a metal electrode connected to the graphene stack.

\begin{figure*}[t]
    \centering
    \includegraphics[width=0.8\textwidth]{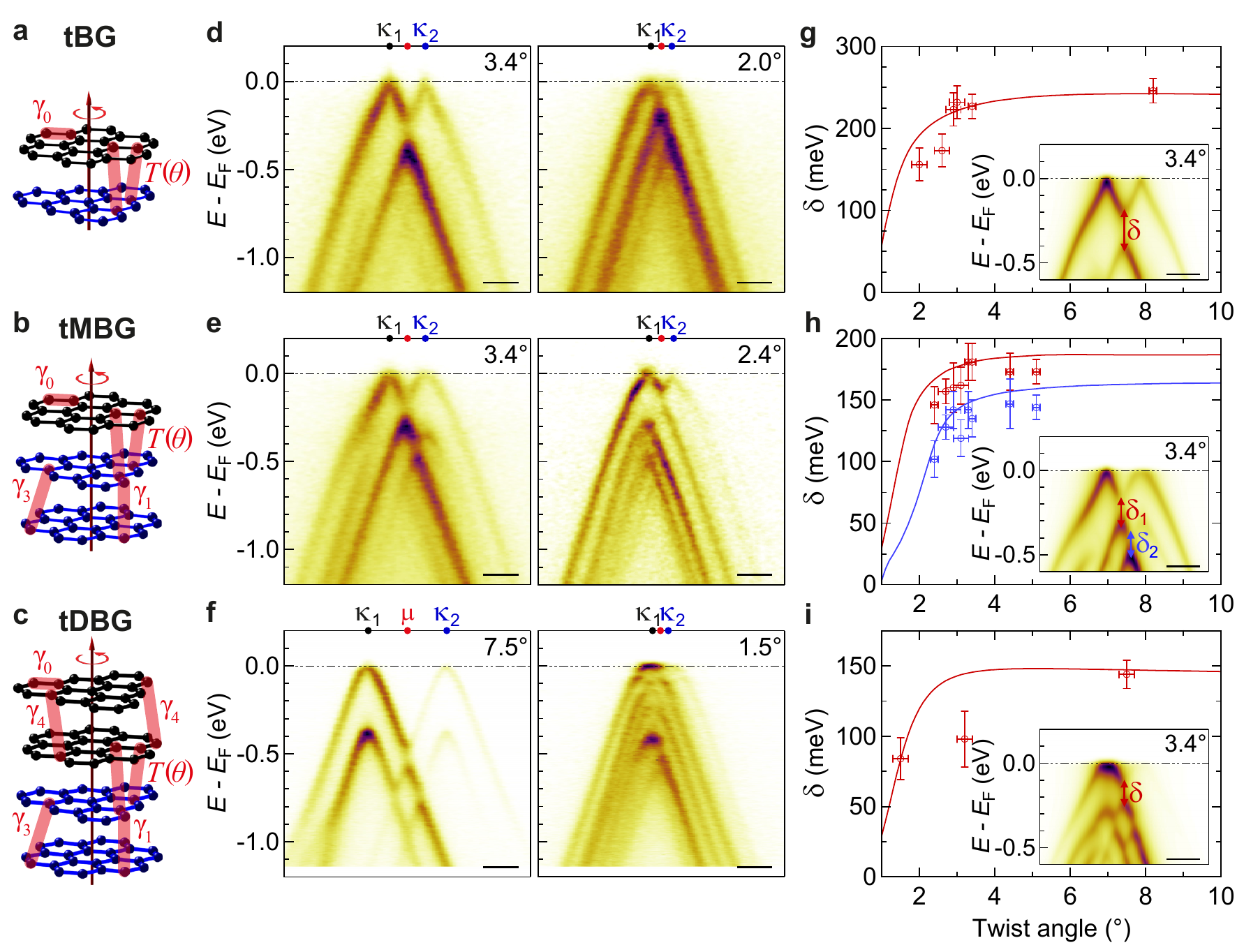}
    \caption{\textbf{Twist angle and layer number dependence of ARPES energy momentum spectra.} \textbf{a-c}~Schematics of tBG, tMBG and tDBG, respectively, labelled with the inter- and intra-layer coupling parameters. \textbf{d-f}~Experimental ARPES energy-momentum cuts in the $\upkappa_{1}$-$\upkappa_{2}$ direction for tBG, tMBG and tDBG, respectively, at 2 twist angles. \textbf{g-i}~Plots of hybridisation gap size vs twist angle for tBG, tMBG and tDBG, respectively. The solid lines correspond to the data extracted from the simulated spectra, and the data points to those from the experimental spectra. Insets show simulated ARPES spectra for the corresponding layer geometries, at $\theta =$~3.4$^\circ$, with the hybridization gaps labelled. All scale bars are 0.1~\AA$^{-1}$. The first panel in \textbf{e} is from monolayer on bilayer graphene, the second is from bilayer on monolayer graphene.}
    \label{figtwist}
\end{figure*}

A hybrid $\textbf{k}\cdot\textbf{p}$ theory-tight-binding model was used to calculate the electronic structure of twisted few layer graphenes, using a full set of Slonczewski-Weiss-McClure (SWMcC) parameters for the aligned multi-layer graphenes~\cite{SW, MCC1, MCC2}. The values of the SWMcC parameters, as given in Table~1 of Methods, were established by earlier transport studies~\cite{Yin2019} and are used here without fitting to the experimental data. Details of the calculations are given in Methods and SI section~3. We focus on the electronic band structure across the moiré Brillouin zones (mBZs), at the graphene Brillouin zone corners, Fig.~\ref{fig1}b. Band-folding results in a rich electronic band structure, as shown in Fig.~\ref{fig1}c for twisted monolayer-on-bilayer graphene (tMBG) at a twist angle of $\theta = 3.4^\circ$.

Not all of these bands are apparent in ARPES spectra and their relative intensities change with measurement conditions. The photoemission intensity depends on matrix elements for the photoexcitation process, resonance and interference effects, and attenuation of the photo-emitted electrons~\cite{Damascelli2004}. For complex systems, this can lead to confusion over the interpretation of the ARPES spectra and their relation to electronic band structure calculations, necessitating simulation of the ARPES intensity. To do this, the probability of a photo-stimulated transition from an initial band state in graphene to a hybrid $\textbf{k} \cdot \textbf{p}$ state in vacuum was calculated using Fermi’s Golden Rule~\cite{Zhu2021,Thompson_2020}, as described in Methods and SI section~4. The final state ($\psi_\text{vac}$) was assumed to be a plane wave in the vacuum. Travel of the photoelectron to the surface, and escape and detection, were included by accounting for an increased path length for emission from the lower layers, resulting in a phase difference in the plane waves and an attenuation in intensity as shown schematically in Fig.~\ref{fig1}d. This phase difference was determined from the out-of-plane component of the final state momentum, $k_z$; the validity of this approach was tested by comparison of simulation to measurement for photon energy dependent spectra of bilayer graphene (see SI section~4). Finally, the simulated spectra were convoluted by a Lorentzian peak of width 60~meV to account for experimental broadening~\cite{MuchaKruczyski2008} due to sample quality, intrinsic linewidth and measurement resolution. 

Simulated ARPES spectra for tMBG at $\theta = 3.4^\circ$, are given in Fig.~\ref{fig1}e and compared to the corresponding experimental spectra in Fig.~\ref{fig1}f, for which the measured twist angle is $\theta=3.4 \pm 0.1^\circ$. The twist-angle was determined from constant energy maps near $E_F$, using the replica bands to determine the mBZ and hence $\theta$. Energy-momentum slices through the corners of the mBZ show the Dirac cones of the primary bands, moiré replica bands, and hybridisation gaps where bands from the rotated monolayer anti-cross with those of the bilayer. Photoemission from bands in the upper monolayer graphene (MLG) is more intense than from those in the bilayer graphene (BLG) underneath, with the replica bands lower in intensity than the corresponding primary bands. 

In the ARPES constant energy maps, plotted over the mBZs (shown in red) in Fig.~\ref{fig1}e, the primary and replica bands are readily identified near $E_\text{F}$, with the Dirac cones at the mBZ $\upkappa$ points. However, at deeper energy cuts, interactions between bands make it harder to assign the origin of the photoemission intensity to a specific band in a given layer. The band decomposition at these constant energy slices, determined from the electronic band structure calculations, is shown for comparison in SI section~5. Despite this complexity, it is clear that the model accurately captures both the relative spectral intensities and band positions of the experimental spectra, enabling the electronic structure to be probed in greater detail.

\textbf{Twist-angle and layer-number dependent spectra.} 
Comparison of spectra acquired from different twist-angles and stacking orders allows a quantitative test of band structure predictions from the ARPES data. In Fig.~\ref{figtwist}, ARPES energy-momentum slices are presented for twisted bilayer graphene (tBG), twisted monolayer-on-bilayer graphene and twisted double bilayer graphene (tDBG). The twist angle is defined relative to Bernal stacking. 

For larger twist angles, $\theta>3^\circ$, the primary bands are readily resolved in the ARPES spectra with only faint replica bands, whilst at smaller twist angles the intensities of the replica bands increase and the electronic band structure becomes more complex. Quantitative analysis of the replica band intensities near $E_\text{F}$ is given in SI section~6 for tBG at different twist angles, showing good agreement between the experimental data and the simulations. In general, the intensity of the replicas decreases in successive mBZs away from the primary bands, as expected due to the lower probability of scattering further in reciprocal space.

Where the bands meet, hybridisation between them results in anti-crossings, with the gaps most obvious at or near the $\mu$ point in the slices shown here.  The size of the gaps ($\updelta$) at the anti-crossings of the primary bands are plotted in Figs.~2g-i as a function of the twist angle. These were determined by fitting of energy distribution curves (EDCs), as illustrated in the insets of Fig.~2 and described in detail in SI section~7. $\updelta$ depends on both the interlayer and intralayer coupling parameters, hence the agreement between experiment (data points) and theory (solid lines) across all twist angles and stacking arrangements demonstrates the accuracy of the theoretical approach for describing the twisted interface. Note that the same SWMcC parameters were used for each structure, without fitting to the experimental data. 

At larger twist angles, $\theta \geq 4^{\circ}$, the simulations predict that $\updelta$ is roughly constant with twist angle, verified by the agreement to the experimental results. In this regime, the anti-crossings occur at energies at which the bands are well described by a linear dispersion and hence $\updelta$ scales only with the strength of the potential that couples the states in the different layers. In these simulations, the magnitude of variation of the moiré potential is a constant factor independent of twist angle, related to the interlayer coupling parameters. However, at smaller $\theta$, the magnitude of the hybridization gap depends sensitively on twist angle and changes subtly with stacking geometry, Figs.~\ref{figtwist}g-i. At these small twist angles, the anti-crossings lie close to the Dirac points and distort the linear dispersion, as can be seen in the ARPES spectra, forming a band whose bandwidth decreases with twist angle.

\begin{figure}[t!]
    \centering
    \includegraphics{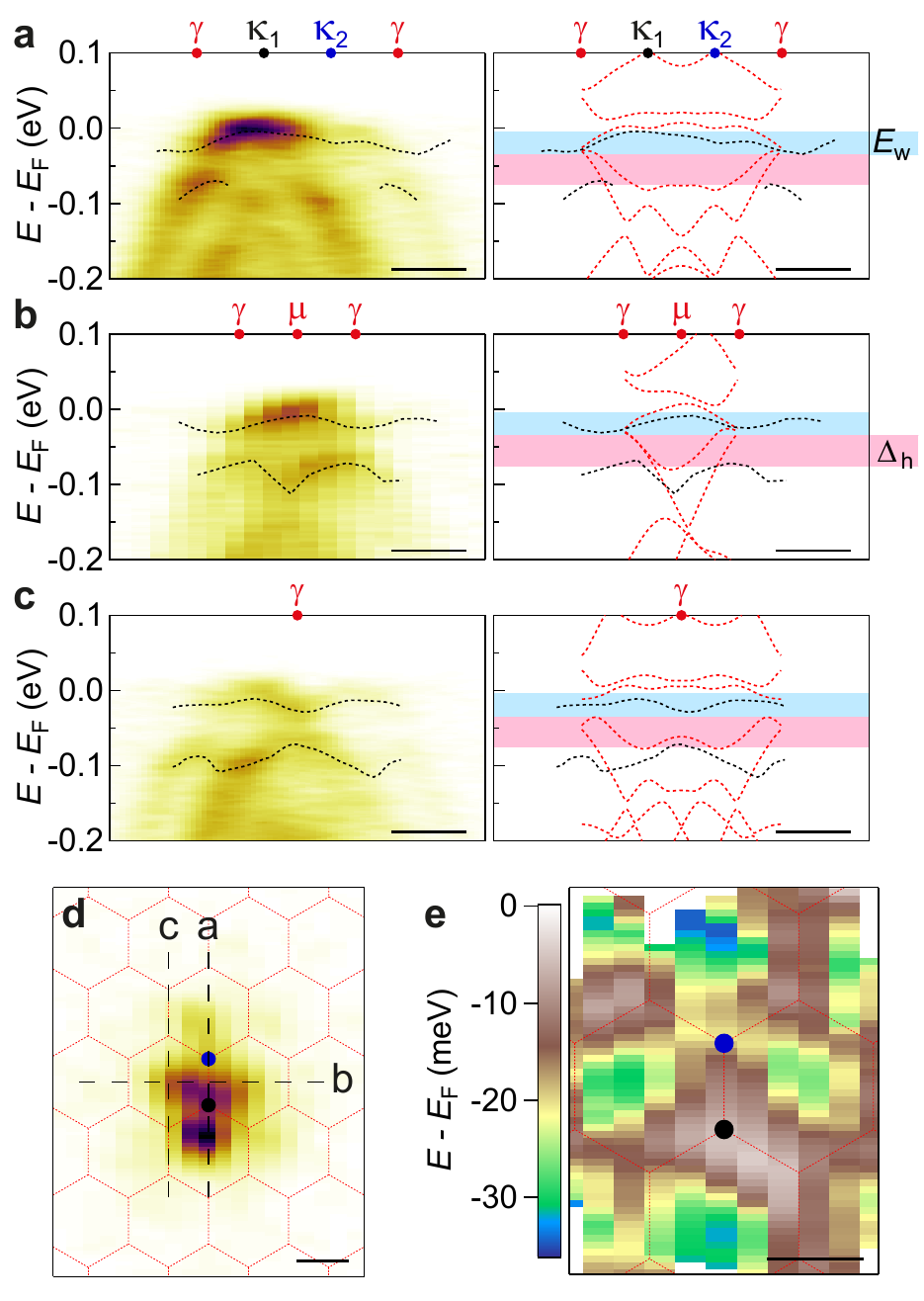}
    \caption{\textbf{Flat band dispersion in 1.5$^\circ$ tDBG.} \textbf{a-c}~Energy-momentum cuts (left-hand) along the black dashed lines in \textbf{d} and the corresponding band dispersions (right-hand). The black lines correspond to the peak positions extracted from the experimental data by fitting EDCs and the red lines corresponding to the predicted electronic structure. \textbf{d}~ARPES constant energy plot at $\textit{E}-\textit{E}_\text{F} = 30$~meV with the mBZs overlaid in red. \textbf{e}~Energy of the flat band plotted in the $\textit{k}_\textit{x}-\textit{k}_\textit{y}$ plane, with the mBZs overlaid in red. All scale bars are 0.05~\AA$^{-1}$.}
\label{figflat}
\end{figure}

\textbf{Flat band analysis in tDBG.}
In the smallest twist angle sample measured here, tDBG at $\theta=$~1.5~$\pm$~0.2$^\circ$, an almost flat valence band is observed at $E_\text{F}$ (Fig.~2f, right-hand panel). A more detailed analysis of this band is shown in Fig.~\ref{figflat}. Energy-momentum cuts through the high symmetry points (Figs.~\ref{figflat}a-c with their directions indicated on the constant energy plot in Figs.~\ref{figflat}d) show ARPES intensity near $E_\text{F}$ in all directions, corresponding to the flat band, with a clear gap to the lower lying valence band states. 

The band dispersion across the first few mBZs was found from the experimental spectra by fitting EDCs and is shown by the black lines in the right-hand panels of Figs.~\ref{figflat}a-c, with the predicted band structure in red. Fig.~\ref{figflat}e shows the energy of this valence band edge as a $k_x-k_y$ plot; the corresponding plot from the simulations is shown in SI section~8. Although it is at the limit of the experimental resolution here, the flat band has a weak dispersion, periodic across the mBZ as required, with the band minimum at $\gamma$, and is clearly gapped from the lower lying valence bands across all of the mBZ. 

\begin{figure*}[t!]
    \centering
    \includegraphics[width=0.9\textwidth]{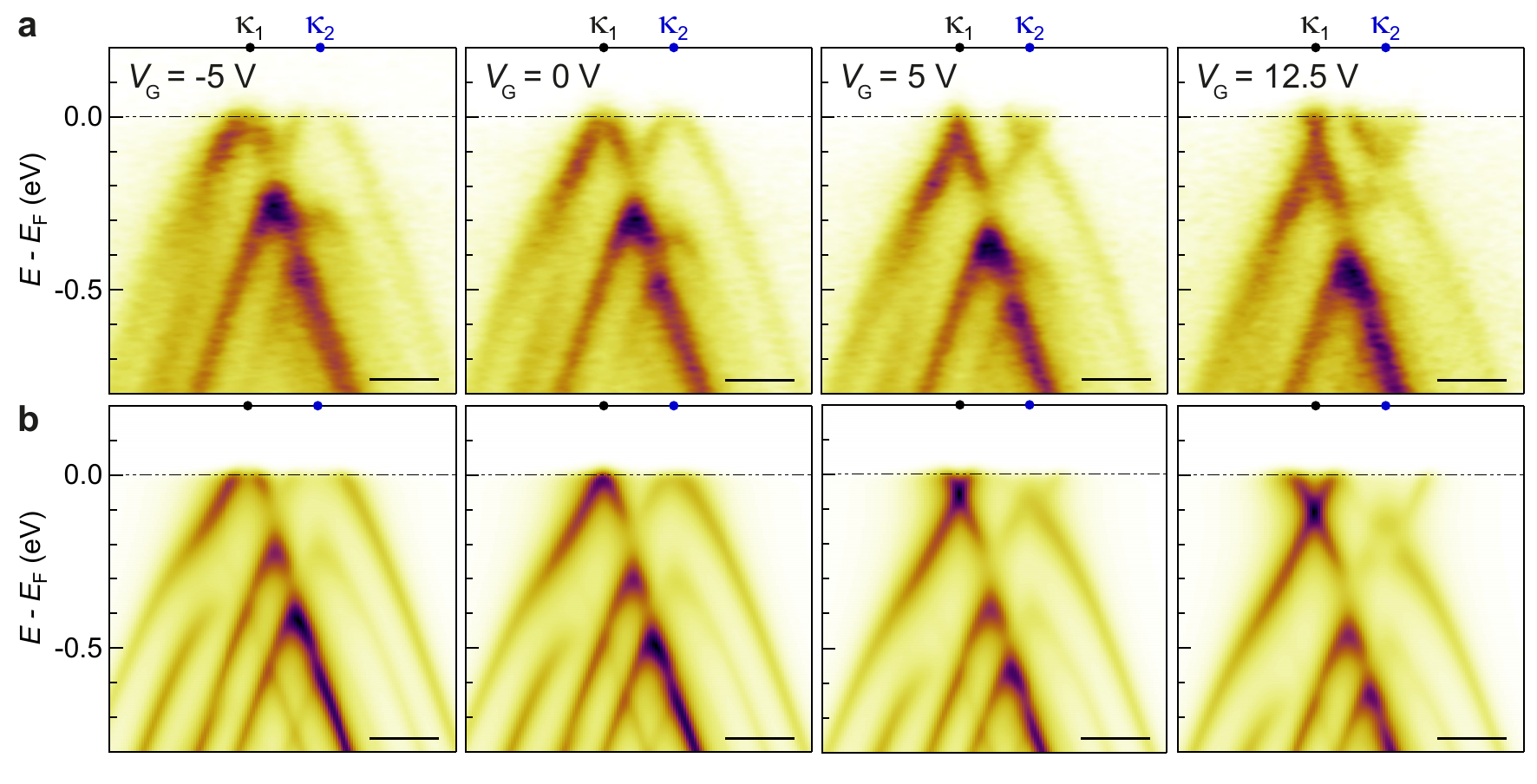}
    \caption{\textbf{Electrostatic gating of tMBG.} \textbf{a}~ARPES energy-momentum cuts of 3.4~$\pm$~0.1$^\circ$ tMBG taken along the $\upkappa_{1}$-$\upkappa_{2}$ direction taken at the labelled gate voltages. \textbf{b}~Simulated ARPES spectra for 3.4$^\circ$ tMBG at varying back-gate voltage, as labelled. All scale bars are 0.1~\AA$^{-1}$.}
    \label{figgate}
\end{figure*}

The key band parameters can be determined from this data. The band-width of the flat valence band at $E_\text{F}$ is measured to be $E_w = 31\pm5$~meV, in good agreement with the predicted value from the electronic structure calculations of $E_w = 33$~meV. The band gap to the next occupied valence band state is smallest at $\gamma$, where it is measured to be $\Delta_h=46\pm5$~meV, significantly greater than the predicted value of $\Delta_h=5$~meV. Note that the electronic structure calculations here do not incorporate the effects of lattice relaxation which are expected to be significant in determining the low-energy electronic structure in  twisted graphenes for small $\theta$, close to or below the magic-angle~\cite{Carr2020,Zhu2021}. For tDBG at $\theta = 1.5^\circ$,  Haddadi et al. found that the gap at $\gamma$ increased by roughly an order of magnitude from $\Delta_h \sim 5$ meV to $\Delta_h \sim 40$ meV when lattice relaxation was included~\cite{Haddadi2020}, consistent with our experimentally determined value. The bandwidth is predicted to decrease further to $E_w \sim 5$~meV at the magic-angle of $\theta = 1.3^\circ$, with the gap staying roughly similar in magnitude. A spectrally isolated flatband such as this has been proposed to be favorable for the emergence of correlated insulators~\cite{Burg2019} and these results prove that, despite previous reports~\cite{Lee2019,Koshino2019},  a vertical displacement field is not required to produce such a band in tDBG.

\begin{figure*}[ht]
    \centering
    \includegraphics[width=0.9\textwidth]{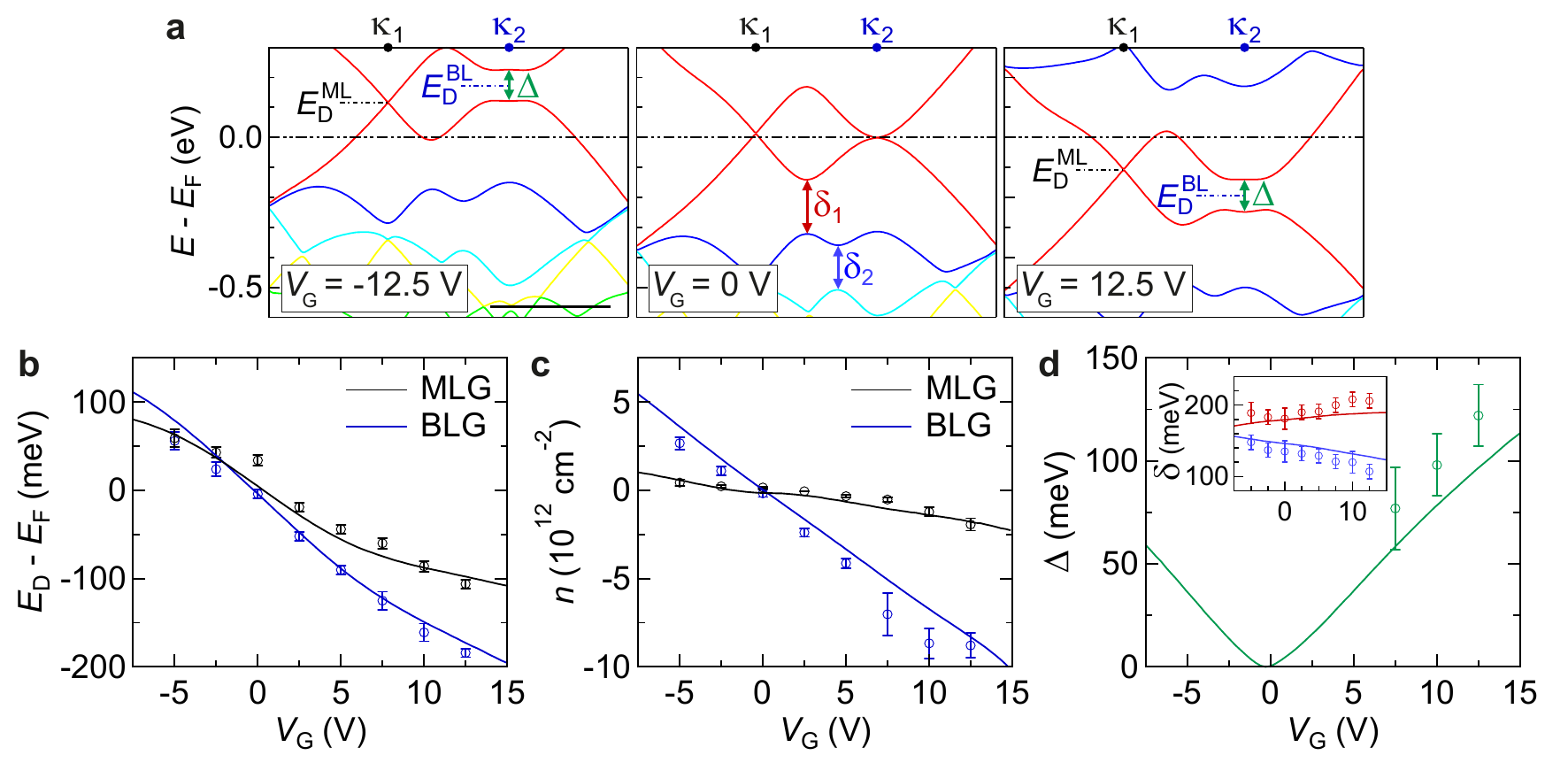}
    \caption{\textbf{Analysis of band structure changes and doping with $\textit{V}_\text{G}$ for 3.4$^\circ$ tMBG, on 26~nm hBN.} \textbf{a}~Band structure around the Fermi level for tMBG at different $\textit{V}_\text{G}$. Labels show the Dirac points ($\textit{E}_\text{{D}}$), bilayer gap size ($\Delta$) and hybridisation gap sizes ($\updelta$). Scale bar is 0.1~\AA$^{-1}$. \textbf{b,c}~Dirac point energies, $\textit{E}_\text{D}$,  and carrier densities, $\textit{n}$, respectively, as a function of $\textit{V}_\text{G}$ for the monolayer (black) and bilayer (blue) Dirac cones. \textbf{d}~The energy gap, $\Delta$, at the bilayer Dirac point as a function of $\textit{V}_\text{G}$. Inset shows the hybridisation gaps, $\updelta_1$ in red and $\updelta_2$ in blue, as a function of $\textit{V}_\text{G}$. Data points are experimental values extracted from the ARPES data, while solid lines are extracted from the simulations. }
    \label{figgatedparam}
\end{figure*}

\textbf{Back-gate-dependent electronic structure.}
Integrating a back-gate electrode into the tMBG heterostructure, as shown schematically in Fig.~1a, allows the gate dependence of the electronic band structure to be investigated. ARPES spectra at varying $V_\text{G}$ for a 3.4~$\pm$~0.1$^\circ$ tMBG sample, with hBN dielectric thickness of $d=26$~nm, are shown in Fig.~\ref{figgate}a. For positive applied back-gate voltage, $V_\text{G}>0$, the Dirac point energies move below $E_\text{F}$, corresponding to $n$-doping. There is no apparent broadening of the spectra, indicating a uniform applied field. Consistent with it being the lower layer, closer to the gate electrode, the Dirac point of the BLG, at momentum $\upkappa_{2}$ and energy $E_\text{D}^\text{BL}$, shifts more than that of the MLG, at momentum $\upkappa_{1}$ and energy $E_\text{D}^\text{ML}$. This indicates partial screening of the back gate~\cite{Slizovskiy2021} and a displacement field across the twisted graphene layers which also opens an energy gap, $\Delta$, at the Dirac point of the BLG~\cite{Zhang2009,McCann2006}. The shift of the BLG bands relative to those of the MLG means that the band anti-crossings occur at slightly different energy and momenta, subtly changing the interactions between bands and hybridisation between layers.

We incorporate the effect of electrostatic gating into the simulations using a self-consistent analysis of on-layer potentials~\cite{Slizovskiy2021}. Starting from the band structure without applied field, potential differences between the layers are added that are proportional to the applied field. The charge redistribution across the layers is determined and the charge density in each layer is used to calculate the screening fields and the resultant modified interlayer potentials. The band structure is recalculated using these modified interlayer potentials and the process iterated to convergence to give a self-consistent response to the applied $V_\text{G}$ (see SI section~9 for further details). Simulated ARPES spectra at varying $V_\text{G}$ are shown in Fig.~\ref{figgate}b, for the same sample geometry as the experimental data (tMBG at $\theta=3.4^\circ$ with a hBN dielectric thickness of $d=26$~nm, and hBN dielectric constant $\epsilon_\text{hBN}=4$~\cite{hbn}).

Changes to the band dispersion with $V_\text{G}$ are shown in Fig.~\ref{figgatedparam}a, emphasising the dominant effects of the applied field: $E_\text{D}^\text{BL}$ and $E_\text{D}^\text{ML}$ shift relative to $E_\text{F}$, indicating electrostatic doping in both layers; $E_\text{D}^\text{BL}$ shifts relative to $E_\text{D}^\text{ML}$, consistent with a field transverse to the layers; this field opens a gap, $\Delta$, at the Dirac point of the bilayer graphene~\cite{McCann2006,Zhang2009}. These key parameters were determined by fitting of the spectra. The change in Dirac point energies is plotted in Fig.~\ref{figgatedparam}b (solid line from the simulations, data points from the experiment, blue corresponds to BLG and black to MLG). For $V_\text{G}>0$,  $E_\text{D}^\text{BL}-E_\text{F}<E_\text{D}^\text{ML}-E_\text{F}<0$, corresponding to electron-doping, while for $V_\text{G}<0$, $E_\text{D}^\text{BL}-E_\text{F}>E_\text{D}^\text{ML}-E_\text{F}>0$ corresponding to hole-doping. The resultant charge density, $n$, is plotted in Fig.~\ref{figgatedparam}c. For the simulations, $n$ is calculated directly from the electronic band structure models by counting the charge in each layer, the experimental data are calculated from $E_\text{D}^\text{BL}$ and $E_\text{D}^\text{ML}$ as described in SI section~10. 

The charge densities in the bilayer, $n_\text{BLG}$, and the monolayer, $n_\text{MLG}$, both scale roughly linearly with $V_\text{G}$, but at a lower rate in the monolayer, $n_\text{MLG} \sim  n_\text{BLG}/4$, such that most of the charge is localised in the BLG, screening the MLG from the gate. At $V_\text{G} = 15$~V there is almost 100~meV difference between $E_\text{D}^\text{ML}$ and $E_\text{D}^\text{BL}$, corresponding to a strong Stark shift due to the displacement field. This changes the BLG dispersion, opening a band gap that scales roughly linearly with the magnitude of $V_\text{G}$ and at $V_\text{G} = 15$~V is again of order 100~meV. Finally, we note that the shift of the BLG bands relative to those of the MLG results in subtle changes to the interlayer coupling, as can be seen through analysis of the hybridisation gaps. For example, the gaps at the anti-crossings of the primary bands, $\updelta_1$ and $\updelta_2$ as labelled on the band dispersion in Fig.~\ref{figgatedparam}a central panel, change in opposite direction with $V_\text{G}$, as shown in the inset of Fig.~\ref{figgatedparam}d. For all band parameters, there is good agreement between the experimental measurements and the simulations, confirming the validity of the model used.


We note that our results illustrate a challenge to applying ARPES with \textit{in-situ} gating to the study of, for example, filling-factor dependent band renormalization in MATBG: a back-gate does not just tune the Fermi level, it also applies a transverse field that subtly changes the hybridisation between layers and hence the Fermi surface. Despite this, ARPES offers a unique capability for resolving the layer-dependent electronic band structure in 2D heterostructures and the evolution of this structure with applied electric field, a crucial control parameter for 2D devices. 

\textbf{Conclusions.} Through comparison between measured and simulated ARPES spectra, we have tested the validity of the hybrid $\textbf{k}\cdot\textbf{p}$ theory-tight-binding model for predicting the electronic band structure of twisted few layer graphenes in the small twist angle regime. The simulated spectra quantitatively agree with the measurements, not only from their band dispersions but also their spectral weights, across a range of twist angles ($1^\circ < \theta <8^\circ$) and numbers of layers (tBG, tMBG, tDBG) with a single set of empirically derived parameters describing the inter and intra-layer coupling for twisted and aligned layers. Detailed analysis of the flat band dispersion in twisted double bilayer graphene at $\theta=1.5\pm0.2^\circ$, close to the magic angle of $1.3^\circ$~\cite{Haddadi2020}, shows that although there is close agreement between the hybrid $\textbf{k} \cdot \textbf{p}$ model and the experimental data for the width of the valence band at the Fermi energy, at $E_w\sim30$ meV, the gap to the lower lying valence band states is larger than predicted, $\Delta_h=$46~$\pm$~5~meV, consistent with the importance of lattice relaxation effects at twist angles close to the magic angle. ARPES with \textit{in-situ} gating reveals the evolution of the electronic band structure with the application of a back gate electrode, demonstrating the importance of both doping and transverse electric field, with quantitative agreement to predicted spectra achieved through a self-consistent approach to modelling the electronic band structure changes with gating. The results reinforce the importance of Stark shifts in 2D heterostructures, even for metallic 2D materials. With this validation, the models can be used with confidence to explore the electronic band structure and emergent transport and optical properties of twisted few-layer graphenes.

\vspace{3mm}
\noindent \textbf{Acknowledgements}
We thank Diamond Light Source for beamtime (proposals SI20573, SI28919 and SI32737).RG acknowledges support from the Royal Society, ERC Consolidator grant QTWIST (101001515) and EPSRC grants EP/V007033/1, EP/S030719/1 and EP/V026496/1. NW acknowledges support from EPSRC grant EP/T027207/1. VF acknowledges support from European Graphene Flagship Core3 Project and EPSRC Grants P/W006502/1 and EP/S030719/1. AM was supported by the EPSRC CDT Graphene-NOWNANO and JN by the University of Warwick and Diamond Light Source.
\vspace{3mm}

\noindent \textbf{Methods}
\vspace{1mm}

\noindent \textbf{Sample fabrication}

Samples were fabricated in an argon atmosphere using a modified PMMA-based tear-and-stack technique~\cite{Frisenda2018}, controlled by a remote micromanipulation rig. Pre-prepared hBN on graphite flakes (on SiO$_2$ substrates) were used as an adhesive layer to tear the graphene flakes supported on a PMMA membrane, allowing accurate control over the twist angle. Samples were annealed in UHV at 300$^{\circ}$C for several hours prior to measurement. For further details see SI section~1.

\vspace{3mm}
\noindent \textbf{Angle resolved photoemission spectroscopy}

ARPES experiments were performed at the nanoARPES branch of the I05 beamline of Diamond Light Source. A choice of two focusing optics are available to perform spatially resolved ARPES: a Fresnel zone plate for submicrometre spatial resolution, and a capillary mirror for improved flux and energy resolution ($\sim$\SI{4}{\micro\metre} spatial resolution). All photoemission spectra in the main text were measured using the capillary mirror, apart from the tMBG data in the second panel of Fig.~2e which was measured with the zone plate, with a 90~eV photon energy and linearly polarised light, at a sample temperature of $< 85$~K. Experimental constant energy cuts were averaged over $\pm$5~meV of the stated energy. For further details see SI section~2.

\vspace{3mm}
\noindent \textbf{Tight-binding modelling of electronic structure}
 A hybrid $\textbf{k} \cdot \textbf{p}$ theory-tight-binding Hamiltonian was used for the twisted structures, as previously reported~\cite{Bistritzer2011, GarciaRuiz2021, Xu2021}. The SWMcC parameters used for this description are shown in Table 1 and are taken from~\cite{Yin2019}. Further details are given in SI section~3.

\begin{table}[H]
\centering
\begin{tabular}{ c c c c c} 
         \hline \hline
        $\gamma_0$ (eV)& $\gamma_1 $ (eV) & $\gamma_3$ (eV) & $\gamma_4$ (eV) &  $\Delta^{\prime} (eV) $  \\
        3.16    &   0.39    & 0.315 &   0.07    &   0.025 \\
        \hline \hline
\end{tabular}
\caption{SWMcC parameters, as in~\cite{Yin2019}.}
\end{table}

\textbf{ARPES simulations.}
The general form of the ARPES intensity in the central mBZ is written as:
\begin{equation}
    \begin{split}
                I  &\propto  | \left<\psi_\text{vac} | \nabla _\textbf{k}H | \psi_\text{band} \right>|^2 \mathcal{L}(E_\textbf{p} +W - \epsilon_\textbf{q} - \hbar \omega) \\
                    & \propto|\sum_{l=1}^3(c_{l,A}e^{i \theta_A}+c_{l,B}e^{i \theta_B})*F^{l-1}|^2 \mathcal{L}(E_\textbf{p} +W - \epsilon_\textbf{q} - \hbar \omega),
    \end{split}
\end{equation}
where $H$ is the twisted graphene Hamiltonian~\cite{Shirley1995} (see SI section~3).The initial state, $\psi_{band}$, is the wavefunction of the graphene electron, written as a linear combination of Bloch functions at a given point in momentum space, comprised of layer and sublattice components that are coupled using SWMcC parameters as well as mixed by the moiré \cite{PhysRevB.93.085409}. The components are solved by diagonalising the respective system Hamiltonian and solving for its wavefunctions. From this, the weights ($c_{l,\lambda}$) are computed for each lattice site, where $l$ is the layer number and $\lambda$ is the sub-lattice index. The final state, $\psi_\text{vac}$, is assumed to be a plane wave in the vacuum.  The interaction term $\nabla_\textbf{k} H$ adds a small phase shift to the ARPES spectra. Attenuation and interference after photoemission are accounted for by the term:
$F=Ae^{ik_z \cdot d}$, where $A=0.4$ per graphene layer was determined by comparison to experiment. $d=3.35$~{\AA} is the distance between adjacent layers and $k_z$ is the out-of-plane component of the final state momentum, calculated by determining $k_z$ from the conservation of energy ($E_p+W=\epsilon_q-\hbar\omega$) and the kinetic energy of an emitted photoelectron ($E_p=\hbar^2(k_z^2+k_{\parallel}^2)/2m_e$).  The Lorentzian factor $ \mathcal{L}(E_\textbf{p} +W - \epsilon_\textbf{q} - \hbar \omega)$ broadens the spectra to match the experimental broadening of 60 meV. Further details are given in SI section~4.

\printbibliography
\end{document}


\maketitle

\section{Sample fabrication}

\begin{figure}[H]
    \centering
    \includegraphics[width=0.8\textwidth]{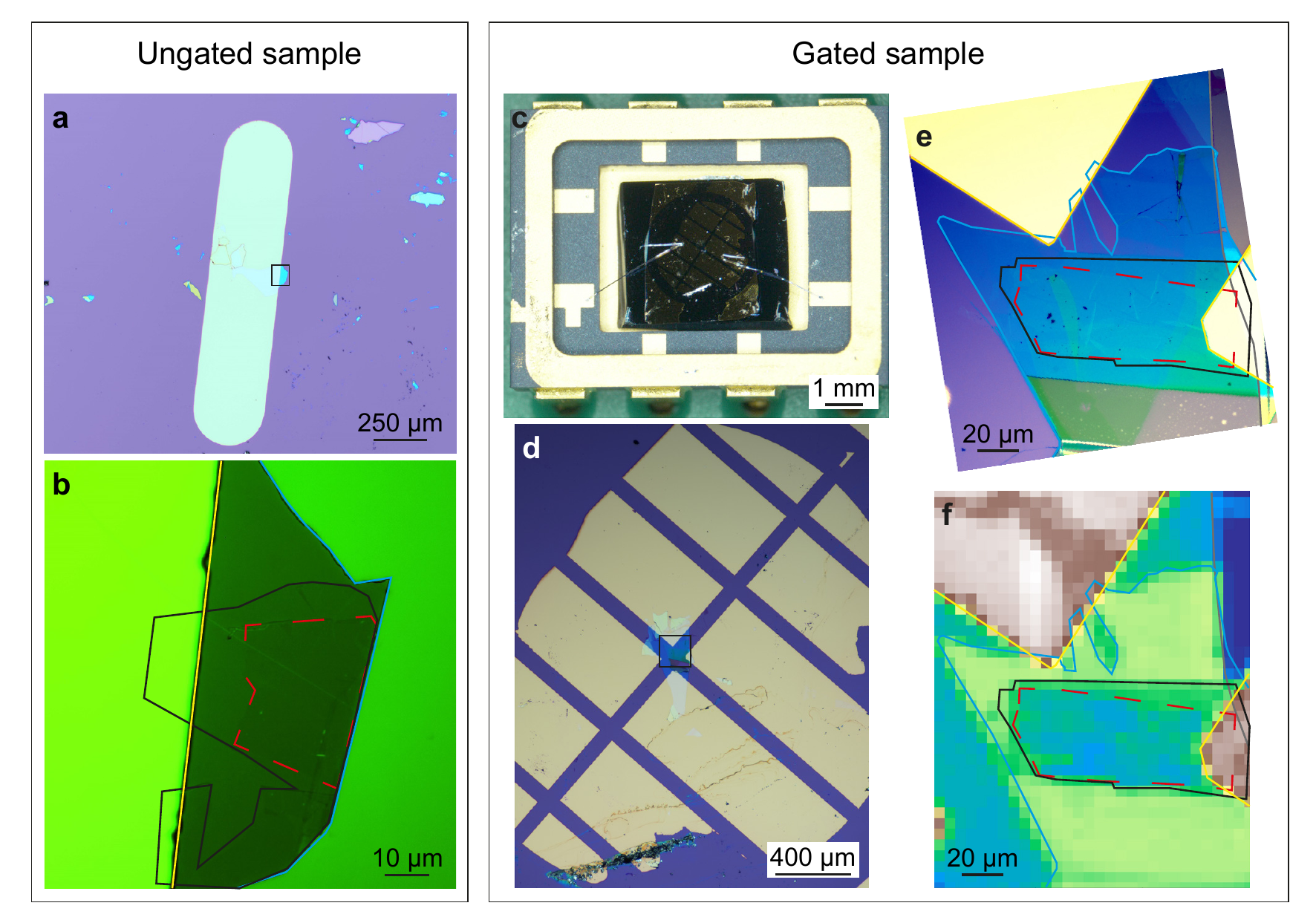}
    \caption{\textbf{Sample summary}. \textbf{a}~Low magnification optical microscope image of a sample with a single, grounding contact (ungated sample). The region of interest (black box) is contacted by a large Au contact for grounding. \textbf{b}~Higher magnification optical microscope image from the black box region in \textbf{a}. A green filter has been used to enhance the image contrast. The coloured outlines mark the Au contact (yellow), hBN (blue) and graphene (black) regions. The red dashed line marks the twisted graphene region. \textbf{c}~Photograph of a chip carrier mounted with a sample with separate electrical contacts to graphene and graphite back-gate (gated sample), with wire bonds between sample contacts and chip carrier pads. \textbf{d}~Low magnification microscope image of the gated sample in \textbf{c}. \textbf{e}~Higher magnification microscope image from the black box region in \textbf{d}. \textbf{f}~SPEM image of the sample region in \textbf{e}. The graphene (black), Au contacts (yellow), hBN (blue) and bottom graphite electrode (grey) sections can all be easily distinguished from each other within the SPEM image, allowing mapping of the sample while in the ARPES chamber.}
    \label{Sample}
\end{figure}

Sample fabrication was performed using a remotely controlled micromanipulation rig housed inside an argon atmosphere. Twisted multilayers of graphene (including tBG, tMBG and tDBG) were transferred onto hBN/Si(SiO$_2$-290nm) and hBN/graphite/Si(SiO$_2$-290nm) stacks for ungated and gated samples respectively. The hBN/graphite heterostructures were prepared using a standard PMMA-based dry transfer technique~\cite{Frisenda2018}, including mechanical exfoliation of crystals on silicon coated with a sacrificial poly-vinyl alcohol (PVA) layer as well as a poly-methyl methacrylate (PMMA) carrier layer. For the twisted graphene multilayers, we employed a modified tear-and-stack technique~\cite{Kim2016} to the PMMA-based dry transfer method to manipulate the twist angle between the graphene layers. The graphene was placed over the edge of the hBN (with the other half touching the SiO$_2$/Si and avoiding the graphite for the gated samples) to tear the graphene layer in half. The sample was then rotated by a target twist angle, $\theta$, before stacking the 2nd half of the graphene layer on top of the first. A single flake of monolayer or bilayer graphene was used for tBG and tDBG twisted samples, respectively. For tMBG twisted samples, flakes consisting of a monolayer attached to bilayer regions were used. The top twisted graphene layer (and bottom graphite flake for gated samples) were contacted using Ti (3 nm)/Au (40 nm) electrodes deposited through a TEM grid shadow mask to minimise contamination. Optical images of ungated and gated samples are shown in Fig.~\ref{Sample}. Gated samples were mounted into chip carriers (Fig.~\ref{Sample}c) using a room temperature curing, UHV compatible, non-conductive epoxy purchased from Atom Adhesives (AA-bond 2116). Electrical connections between the chip carrier and sample electrodes were made using a wire bonder.

Samples were transferred to the beamline in air. Prior to measurement, samples were annealed in UHV at 300$^{\circ}$C for several hours to remove surface adsorbates. Samples without gate electrodes were annealed for $\sim$3 hours, while samples mounted in the chip carriers were annealed for at least 6 hours.

\section{$\mu$ARPES}
ARPES experiments were performed at the nanoARPES branch of the I05 beamline of Diamond Light Source. Here, synchrotron light is focused to a small spot size using specialised optics to allow for high spatial resolution ARPES. A choice of two focusing optics are available: a Fresnel zone plate for submicrometre spatial resolution, and a capillary mirror for improved flux and energy resolution ($\sim$\SI{4}{\micro\metre} spatial resolution). All photoemission spectra in the main text were measured using the capillary mirror (apart from the bilayer on monolayer graphene data in the second panel of Fig.~2e) with a 90~eV photon energy and linear horizontal polarised light. Data was collected using a Scienta Omicron DA30 hemispherical analyser, with a chamber pressure of 1~x~10$^{-10}$~mbar and sample temperature $<$85~K. 

The sample was located and mapped in the ARPES chamber using scanning photoemission microscopy (SPEM). A 4D dataset is collected by raster scanning the sample under the focused beam and collecting an ($E_k,\varphi$) spectrum at each position, where $E_k$ is the photoelectron kinetic energy and $\varphi$ is the emission angle of the photoelectron as measured by the analyser. Each pixel of the SPEM image shown in Fig.~\ref{Sample}f is an integrated intensity of the ($E_k,\varphi$) spectrum from that point on the sample. By comparing this with optical images collected during fabrication, different regions on the sample can be easily identified. $k$-space mapping was performed by rotating the chamber, including the analyser, around the fixed sample and optic, to measure from different polar emission angles. Electrostatic gating measurements were performed using a Keithley~2634B SourceMeter, applying a voltage to the graphite back gate layer, while a separate ground contact connects to the twisted graphene layers adjacent to the measurement position (e.g. Fig S1e). Achieving an effective gate \textit{in-situ} requires good electrical isolation between the two contacts, requiring precise alignment of the contact pads relative to the flakes as well as careful handling. It was found that not all the samples fabricated with gates could be electrically biased \textit{in-situ}.   

\section{Continuum model for twisted few-layer graphene}
For bilayer graphene (BLG), the standard Hamiltonian as defined by McCann et al. \cite{McCann2013} was used. For each twisted structure, a hybrid $\textbf{k} \cdot \textbf{p}$-tight binding Hamiltonian was used for the twisted structures, similar to those previously reported \cite{Bistritzer2011, GarciaRuiz2021, Xu2021}. Such Hamiltonians consider the dispersion around a given valley in reciprocal space and information can be obtained about the system by solving for the eigenfunctions and eigenvalues of the system. The Hamiltonians for twisted bilayer graphene (tBG), twisted monolayer-bilayer graphene (tMBG) and twisted double-bilayer graphene (tDBG), respectively are given below:

\begin{equation}
    \centering
    \mathcal{H}_{tBG} = 
    \begin{pmatrix}
                0 & v \hbar \pi^{\dagger}_{\xi,t} & {11} & \mathcal{T}_{12} \\
                v \hbar \pi_{\xi,t} & 0 & \mathcal{T}_{21} & \mathcal{T}_{22} \\
                 \mathcal{T}_{11}^{\dagger} & \mathcal{T}_{21}^{\dagger} &  0 & v \hbar \pi_{\xi,b} \\
                  \mathcal{T}_{12}^{\dagger} & \mathcal{T}_{22}^{\dagger} &v \hbar \pi_{\xi,b} & 0
    \end{pmatrix},
    \label{tBLG_mat}
\end{equation}

\vspace{3mm}

\begin{equation}
    \centering
    \mathcal{H}_{tMBG} = 
    \begin{pmatrix}
                0 & v \hbar \pi^{\dagger}_{\xi,t} & \mathcal{T}_{11} & \mathcal{T}_{12} & 0  & 0\\
                v \hbar \pi_{\xi,t} & 0 & \mathcal{T}_{21} & \mathcal{T}_{22} & 0 & 0\\
                 \mathcal{T}_{11}^{\dagger} & \mathcal{T}_{21}^{\dagger} &  0 & v \hbar \pi_{\xi,b} & -v_4 \hbar \pi _{\xi,b}^{\dagger}  & -v_3 \hbar \pi_{\xi,b}\\
                  \mathcal{T}_{12}^{\dagger} & \mathcal{T}_{22}^{\dagger} &v \hbar \pi_{\xi,b}  & 0 & \gamma_1 & -v_4 \hbar \pi _{\xi,b}^{\dagger}  \\
                   0 & 0 & -v_4 \hbar \pi _{\xi,b} & \gamma_1 & 0 & v \hbar \pi_{\xi,b} \\
                    0 & 0 & -v_3 \hbar \pi_{\xi,b}^{\dagger} & -v_4 \hbar \pi _{\xi,b} & v \hbar \pi_{\xi,b} &0
    \end{pmatrix},
    \label{tTLG_mat}
\end{equation}

\vspace{3mm}

\begin{equation}
    \centering
    \mathcal{H}_{tDBG} = 
    \begin{pmatrix}
                0 & v \hbar \pi^{\dagger}_{\xi,t} & -v_4 \hbar \pi _{\xi,t}^{\dagger}  & -v_3 \hbar \pi_{\xi,t} & 0 & 0 & 0 & 0\\
                v \hbar \pi_{\xi,t} & 0 &  \gamma_1 & -v_4 \hbar \pi _{\xi,t}^{\dagger} & 0 & 0 & 0 & 0\\
                 -v_4 \hbar \pi _{\xi,t} & \gamma_1 &  0 & v \hbar \pi_{\xi,t} &  \mathcal{T}_{11} & \mathcal{T}_{12} & 0 & 0\\
                  -v_3 \hbar \pi_{\xi,t}^{\dagger} & -v_4 \hbar \pi _{\xi,t} &v \hbar \pi_{\xi,t}  & 0 & \mathcal{T}_{21} & \mathcal{T}_{22} &0 & 0  \\
                   0 & 0 &  \mathcal{T}_{11}^{\dagger} & \mathcal{T}_{21}^{\dagger} & 0 & v \hbar \pi_{\xi,b} & -v_4 \hbar \pi _{\xi,b}^{\dagger}  & -v_3 \hbar \pi_{\xi,b} \\
                    0 & 0 & \mathcal{T}_{12}^{\dagger} & \mathcal{T}_{22}^{\dagger} & v \hbar \pi_{\xi,b} &0 & \gamma_1 & -v_4 \hbar \pi _{\xi,b}^{\dagger} \\
                    0 & 0 & 0 & 0 &  -v_4 \hbar \pi _{\xi,b} & \gamma_1 & 0 & v \hbar \pi_{\xi,b} \\
                    0 & 0 & 0 & 0 & -v_3 \hbar \pi_{\xi,b}^{\dagger} & -v_4 \hbar \pi _{\xi,b} & v \hbar \pi_{\xi,b} &0
    \end{pmatrix}.
    \label{tDBLG_mat}
\end{equation}

$\pi_{\xi,t/b}$ is defined as:
\begin{equation}
    \pi_{\xi,t/b}(p) \approx - \frac{\sqrt{3}a}{2 \hbar} ( \xi p_x + i( p_y+K\theta/2)) + \frac{a^2}{8 \hbar^2}(\xi p_x  -i( p_y+K\theta/2))^2, ~~~~~~~ K=\frac{4 \pi}{3a},
\label{eqn_pi}
\end{equation}
where $p_x,p_y$ are the in-plane components of the momentum shifted to be centred around the $K_+$ valley ($p=\hbar k - \hbar \kappa_{+}$, where $\kappa_{+}=(4 \pi/3a, 0)$ ). In Eq.~(\ref{eqn_pi}), a second-order expansion in momentum is used to account for trigonal warping of the band structure at higher energies.
The interlayer coupling matrices $\mathcal{T}_{ij}$ across the twisted interface are defined as:
\begin{equation}
    \begin{split}
        \mathcal{T}_{k',k}= 
        \begin{pmatrix}
                    \mathcal{T}_{11} & \mathcal{T}_{12} \\
                    \mathcal{T}_{21} & \mathcal{T}_{22}
        \end{pmatrix}
        = \frac{\gamma_1}{3} \sum_{j=0}^2
        \begin{pmatrix}
                    1 & e^{i \xi \frac{2 \pi}{3} j} \\
                    e^{-i \xi \frac{2 \pi}{3} j} & 1
        \end{pmatrix}
       e^{-i K_{j} \cdot \textbf{r}}
    \end{split}
\end{equation}
where $K_{(j)}= K[\text{cos}(2\pi j/3), - \text{sin}(2\pi j/3)]$, (9) with j = 0, 1, 2 and $K=4 \pi /3a$

Table \ref{tableSWMcC} gives the values for the Slonczewski-Weiss-McClure (SWMcC) parameters used for all of the calculations here.

\begin{table}[H]
\centering
\begin{tabular}{ c c c c c} 
         \hline \hline
        $ v$ (m/s) & $\gamma_1 $ (eV) & $v_3$ (m/s) & $v_4$ (m/s) & $\Delta^{\prime}$ (eV)   \\
        1.02 x 10$^6$ & 0.39 & 1.02 x 10$^5$ & 2.27 x 10$^4$ & 0.025 \\
        \hline \hline
\end{tabular}
\caption{The SWMcC parameters used here, with values taken from~\cite{Yin2019}.}
\label{tableSWMcC}
\end{table}

$v \propto \gamma_0$ accounts for the intralayer coupling between the A and B sites in the graphene lattices. $\gamma_1$ corresponds to the interlayer coupling between the dimer sites ($A_b$/$B_t$) of the aligned bilayers and is also used to compute the interlayer coupling at the twisted interfaces. $v_3,~v_4 = \sqrt{3}a\gamma_i/2\hbar$ account for the coupling between the non-dimer sites and the coupling between a dimer site and a non-dimer site, respectively. Lastly, $\Delta'$ accounts for the difference in the on-site electron energies of the dimer sites. $v_4$ and $\Delta'$ also have the effect of adding electron-hole asymmetry into the optical response of the system. 

\section{Simulating ARPES intensity}
The ARPES intensity is proportional to the square of the modulus of the transition amplitude between the initial and final states of the system under the perturbation caused by incoming photons \cite{MuchaKruczyski2008, Zhu2021}. We define the initial state of the system as a single Bloch electron at the surface. The general form of a Bloch wave is:
\begin{equation}
    \Phi_j(r)= \sum_j C_j \varphi_{k,j}(\textbf{r}) \approx \sum_j\sum_{\lambda=A,B} (c_{j,\lambda}e^{i\theta_{\lambda}})e^{i\textbf{G}_m\cdot\textbf{r}},
\end{equation}
where $j$ is the band index of the system and $C_j$ is the amplitude of the wavefunction of the system, containing a sublattice phase. The wavefunctions, $\varphi_{k,j}(\textbf{r})$, are calculated by diagonalising the Hamiltonians shown in Eqs.~(\ref{tBLG_mat}),(\ref{tTLG_mat}) and (\ref{tDBLG_mat}). $\textbf{G}_m$ are the moiré reciprocal lattice vectors. The final state of the system, the emitted photoelectron with momentum $\textbf{p}_e$, is treated as a plane wave with approximate form:
\begin{equation}
    \Phi_f(r) \propto \textrm{exp}\left( \frac{i}{\hbar} \textbf{p}_e \cdot \textbf{r} \right) \approx \mathbb{1}.
\end{equation}
The perturbation due to the incoming radiation is treated as a first-order approximation of the standard perturbative Hamiltonian:
\begin{equation}
    \begin{split}
          \mathcal{H}_p &=  -\frac{2}{2m_e}(\textbf{A} \cdot \textbf{v}+ \textbf{v} \cdot \textbf{A})= -\frac{e}{\hbar c}\textbf{A} \cdot \textbf{v} \\
           &= -\frac{e}{\hbar c}\textbf{A} \cdot \nabla_k H,
    \end{split}
\end{equation}
where \textbf{A} is the electromagnetic vector potential of the incoming radiation, $H$ is the Hamiltonian of the graphene system, and \textbf{v} is the electron velocity operator which introduces the interaction Hamiltonian ($H_{int}=\nabla_k H$) for the photoemission process. 

Accounting for multi-layer effects, the general form of the ARPES intensity in the central mini-Brillouin zone can then be written as:
\begin{equation}
    \begin{split}
        &I \propto \left|<\varphi_f |\textbf{A} \cdot \textbf{v}| \varphi_i> \right|^2 \delta(\epsilon_e+W-\epsilon_M-\hbar \omega) \\
        &= \left|\textbf{A} \cdot <\varphi_f | \nabla_k H| \varphi_i> \right|^2 \delta(\epsilon_e+W-\epsilon_M-\hbar \omega) \\
        & \propto|\sum_{l=1}^3(c_{l,A}e^{i \theta_A}+c_{l,B}e^{i \theta_B})*F^{l-1}|^2 \mathcal{L}(\epsilon_e+W-\epsilon_M-\hbar \omega),
    \end{split}
    \label{ARPES_I}
\end{equation}
where $\hbar \omega$ is the energy of the photons used in the experiment, $W=4.6$ eV is the workfunction of graphene~\cite{Song2012}, and $\epsilon_M$ is the energy of the measured electron in the crystal lattice. The delta function comes from treating a single electron in a many-body system using a time-ordered one-electron Green’s function~\cite{Damascelli2004}. To give a qualitative match to the experimental spectra, accounting for instrument resolution and lifetime broadening etc., a Lorentzian with 60~meV broadening is included ($\mathcal{L}(\epsilon_e+W-\epsilon_M-\hbar \omega)$). \textbf{A} is a 2x1 vector whose value can be changed to alter the polarisation of the light in the system. Eq.~(\ref{ARPES_I}) thus gives an analytical approximation to the ARPES intensity.

\subsection{Accounting for attenuation and interference across multi-layers}
We do not attempt a one-step photoemission model~\cite{Damascelli2004}, but instead account for attenuation and interference of the photo-emitted electrons from different layers using a scaling factor, $F$, assuming a plane wave final state. This scaling factor accounts for attenuation (due to the short mean free path of photo-electrons) and interference (due to layer-dependent phase differences of the photo-electrons):  $F=Ae^{i k_z \cdot d}$, where $k_z$ is the out-of-plane momentum of the photo-emitted electrons and $d=3.35$~\AA~is the distance between adjacent graphene layers. The attenuation factor here was set  to be $A=0.4$ per layer, found by comparison to the experimental data. Starting with the kinetic energy of the emitted photo-electron 
\begin{equation}
    E_p= \hbar^2 \frac{k_z^2 + K_{\parallel}^2}{2 m_e},
\end{equation}
using conservation of energy, the out-of-plane momentum is
\begin{equation}
    k_z=\sqrt{\frac{2 m_e}{\hbar^2} \left[ \hbar \omega-W+\epsilon_M -\frac{\hbar^2}{2 m_e}   \left[ \left(\frac{4 \pi}{3a} +p_x \right)^2 +p_y^2   \right] \right]}.
    \label{kz}
\end{equation}
This defines all terms in the factor $F^{l-1}=(Ae^{i k_z \cdot d})^{l-1}$ seen at the end of Eq.~(\ref{ARPES_I}), where $l$ is the layer number counting from the top surface.

\begin{figure}[H]
    \centering
    \includegraphics[width=0.7\textwidth]{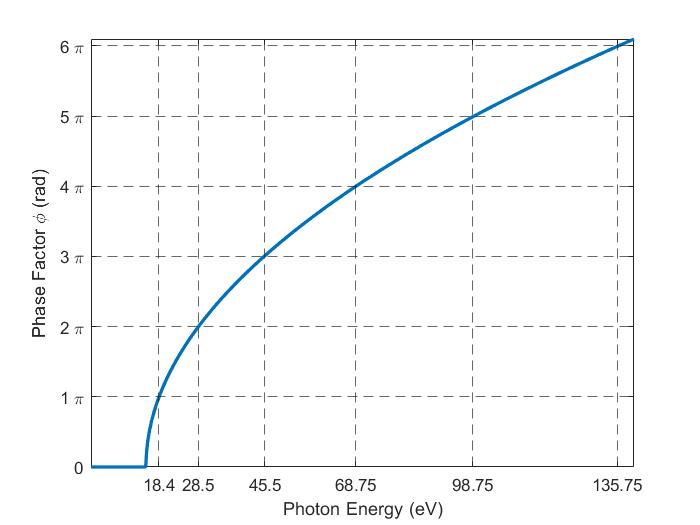}
    \caption{Approximate phase factor, $\phi = \textit{k}_\textit{z} \cdot \textit{d}$, as a function of photon energy.}
    \label{Phase}
\end{figure}

\begin{figure}[H]
    \centering
    \includegraphics[width=0.6\textwidth]{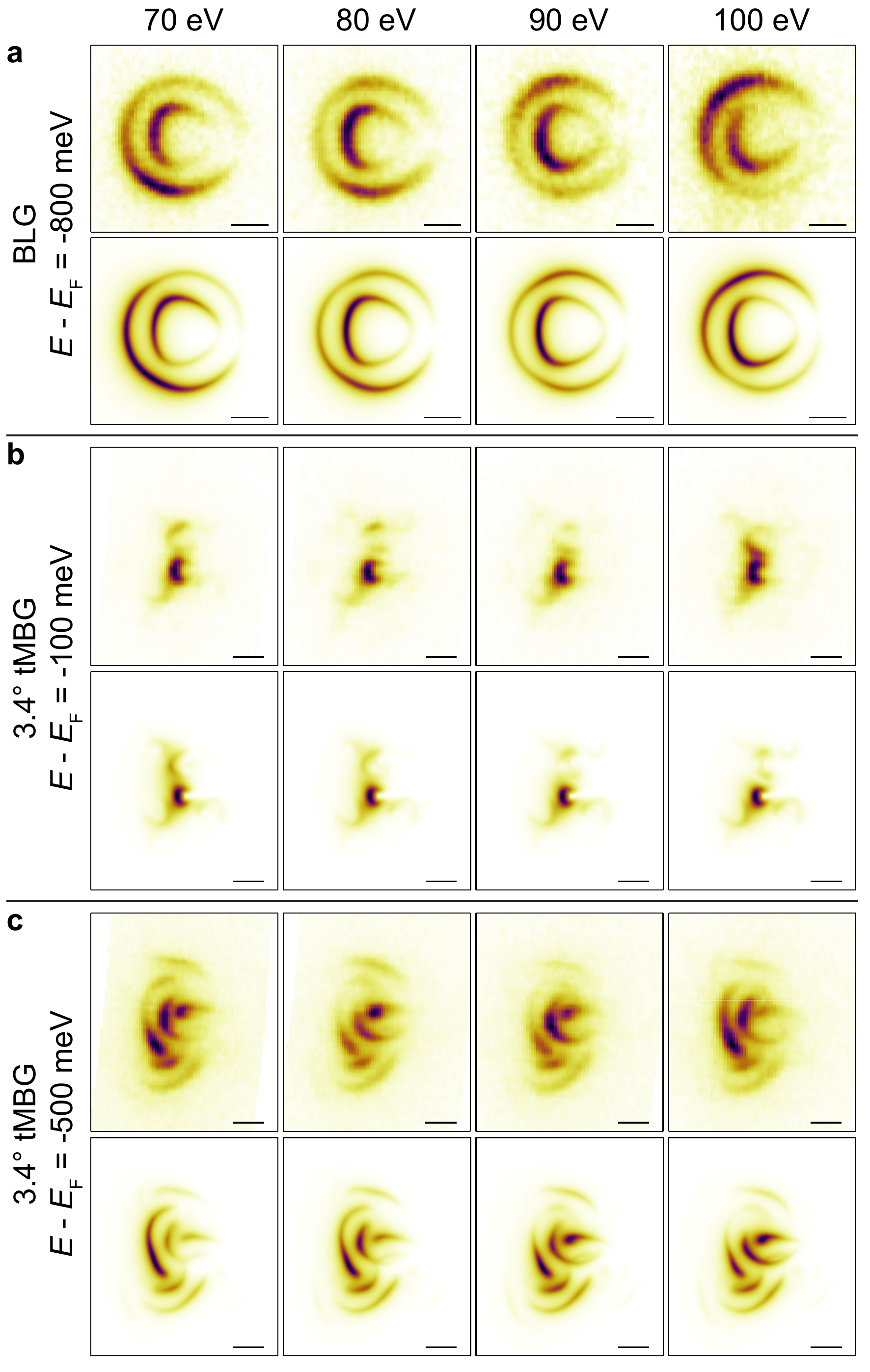}
    \caption{\textbf{Changes in the ARPES spectra of multilayered graphene systems due to varying photon energy}. Constant energy cuts for \textbf{a} bilayer graphene at $\textit{E}-\textit{E}_\text{F}=$~-800~$\pm$~15~meV, \textbf{b} 3.4$^\circ$ tMBG at $\textit{E}-\textit{E}_\text{F}=$ {-100}~$\pm$~5~meV and \textbf{c} 3.4$^\circ$ tMBG at $\textit{E}-\textit{E}_\text{F}=$~-500~$\pm$~5~meV at photon energies (moving left to right) 70~eV, 80~eV, 90~eV and 100~eV. Top panels are from experimental ARPES spectra, while bottom panels are from simulation. All scale bars are 0.1~\AA$^{-1}$.}
    \label{PhotonEnergy}
\end{figure}

As a consequence of Eq.~(\ref{kz}), the phase difference between photoelectrons emitted from different layers depends on the energy of the incident photon used. This phase factor is plotted as a function of the photon energy in Fig.~\ref{Phase}, for photoelectrons emitted from neighbouring graphene layers, across the experimentally relevant photon energies. Note that the phase factor changes significantly with photon energy, indicating that the relative intensity of different features in the ARPES spectra should be strongly dependent on the photon energy. The validity of this approach was tested by comparing experimental and predicted spectra for bilayer graphene and tMBG at varying photon energy, as shown in Fig.~\ref{PhotonEnergy}. While many of the key features are consistent across the different energies, the relative intensities of features changes with photon energy. The results show that although this simple model is not a full description of the photoemission process, it gives a good approximation. We also note that the elongated flat region at the beginning of Fig.~\ref{Phase} corresponds to the fact that it takes $\epsilon=W$ to photoexcite electrons from the $\Gamma$ point but additional energy to photoexcite electrons with nonzero in-plane momentum.

i

We note that the comparison between simulated and experimental spectra in Fig.~\ref{PhotonEnergy} can also be used to confirm the sign of the tight binding parameter $\gamma_1$. There is no clear consensus on the sign of $\gamma_1$ in the literature: in some reports it is assumed to be positive, and in others, negative~\cite{MuchaKruczyski2008}. In other reports, $\gamma_1$ is taken to be negative with a phase factor of $e^{i\pi}$  per layer replacing the factor exp$(ik_z \cdot d)$ in $F$~\cite{Thompson_2020}. A comparison of simulated and experimental spectra demonstrates that the sign of $\gamma_1$ is positive with no additional phase needed. 

\section{Associating spectral features with electronic bands}
\begin{figure}[H]
    \centering
    \includegraphics[width=0.8\textwidth]{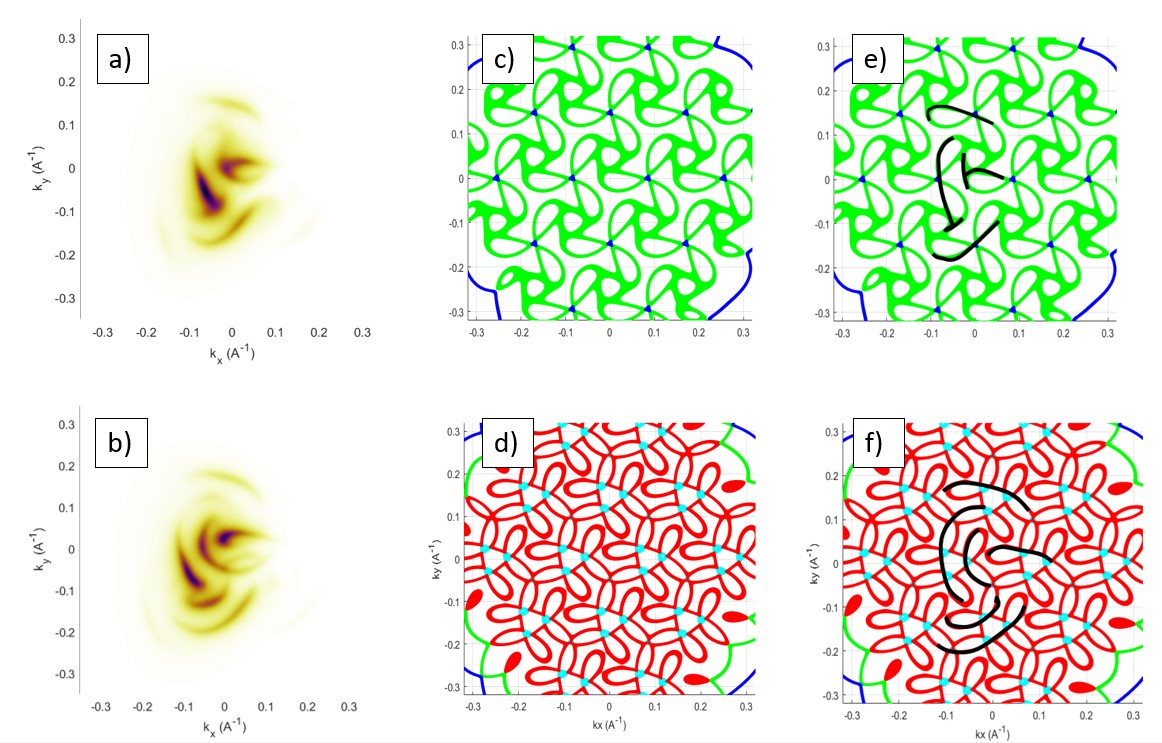}
    \caption{\textbf{Identifying the band contributions in ARPES from tMBG}. Simulated ARPES contant energy cuts at $\textit{E}-\textit{E}_\text{F}=$ \textbf{a} -350~meV and \textbf{b} -500~meV for 3.4$^\circ$ tMBG. \textbf{c,d} band structure of 3.4$^\circ$ tMBG at the same energies as in \textbf{a} and \textbf{b}, respectively. \textbf{e,f} Same as \textbf{c} and \textbf{d} overlaid with black lines to mark the bands contributing to the ARPES intensity in \textbf{a} and \textbf{b}.}
    \label{figdecomp}
\end{figure}

For small-twist angle samples at higher binding energies, the spectra become more complex, as shown in Figs.~\ref{figdecomp}a,b, and it becomes difficult to distinguish which layer and band each spectral feature is associated with. Comparison between the simulated and experimental spectra, alongside the band structure predictions, enables the ARPES spectral features to be assigned to distinct valence bands. For example, in Fig.~\ref{figdecomp}, constant energy cuts of the ARPES spectra are shown alongside the band structure in the $k_x - k_y$ plane averaged over the same energy window, $E~\pm$~30~meV, shown in the extended zone scheme such that the pattern is repeated across successive mBZs. In Figs.~\ref{figdecomp}e,f the bands that are most apparent in the ARPES spectra are highlighted. In the band calculation the colour scheme is as follows: {blue, green, red, cyan} = {1st, 2nd, 3rd, 4th} valence band. 

\section{Analysis of replica band intensity}

\begin{figure}[H]
    \centering
    \includegraphics[width=0.9\textwidth]{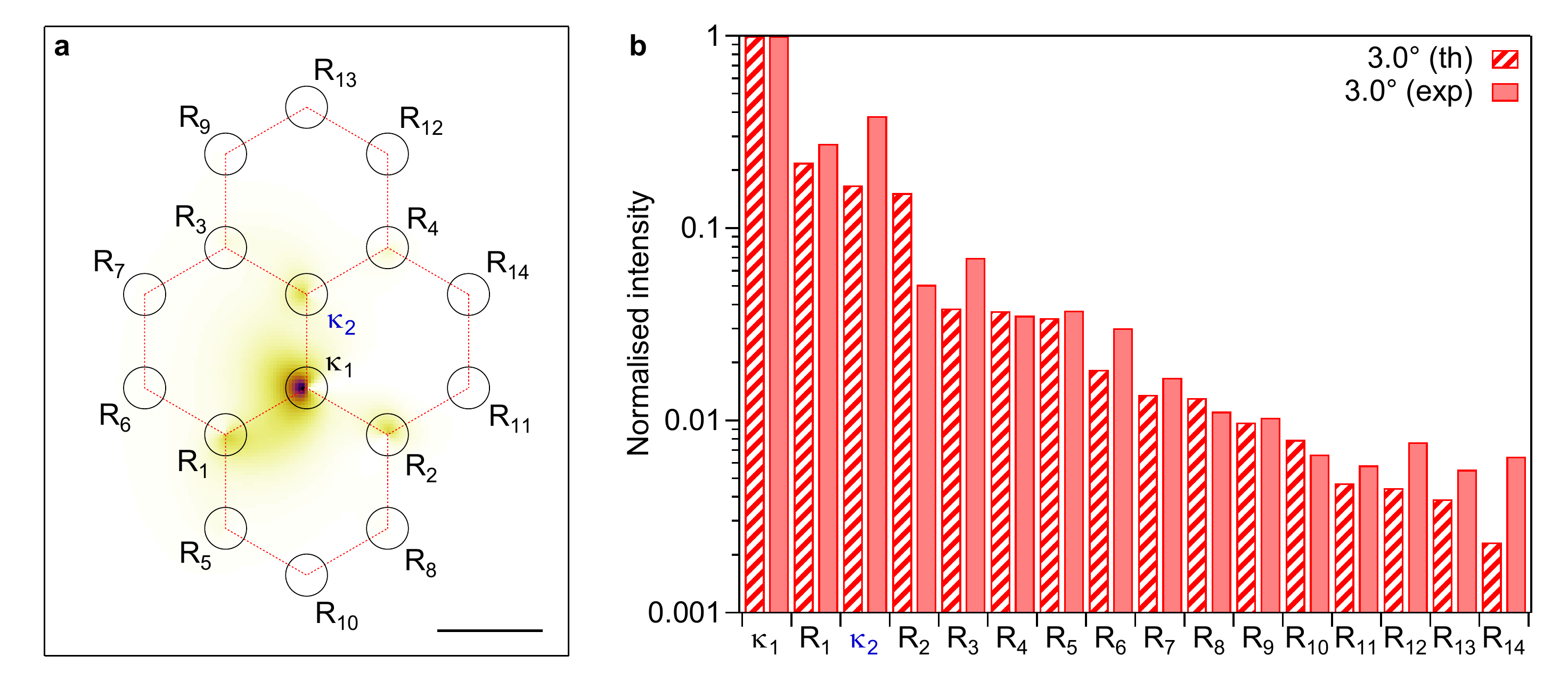}
    \caption{\textbf{Analysis of replica band intensity in 3.0$^\circ$ tBG.} \textbf{a}~Simulated constant energy cut at the Fermi level of 3.0$^\circ$ tBG overlaid with the mBZ and labels for each mBZ corner. Scale bar is 0.1~\AA$^{-1}$. \textbf{b}~Replica intensity of 3.0$^\circ$ tBG from \textbf{a}, normalised by the intensity of the primary band from the top layer, $\upkappa_1$, from experiment and simulation. The contributed intensity comes from the circular areas in \textbf{a}.}
    \label{replica_alt}
\end{figure}

As shown in Section 5, in the extended zone scheme, bands are equivalent in each mBZ. However, the photoemission intensity of the replica bands decreases in successive mBZs due to the reduced probability of scattering to higher wave vectors. Fig.~\ref{replica_alt}a shows the photoemission intensity near the Fermi level for a $3.0^\circ$ tBG. The primary Dirac points are labelled following the convention from the main text of $\upkappa_1$ for the top layer and $\upkappa_2$ for the bottom layer, where these now refer to an intensity. The photoemission intensity from the bottom layer is reduced relative to the top layer due to attenuation, i.e. $\upkappa_2=A\times \upkappa_1$, where $A$ is the previously mentioned attenuation set to match the experiment as $A=0.4$. Intensity due to replica bands can be seen at the corners of the mBZs. These are labelled by the order of their respective intensity, i.e. R$_1$ is the most intense replica and R$_{14}$ is the least intense. This is shown more clearly in Fig.~\ref{replica_alt}b, where the total intensity is taken from the circular regions marked in Fig.~\ref{replica_alt}a, and normalised with respect to the intensity at $\upkappa_1$. This is compared to the experimental results where there is relatively good agreement and they roughly follow the same hierarchy.

The intensity ordering of the replicas is nontrivial. For example, we can associate replicas R$_1$ and R$_2$ with states from the bottom layer, and replicas R$_3$ and R$_4$ with states from the top layer. We know this because, in the case of tMBG, R$_1$ and R$_2$ show a bilayer graphene-like dispersion, while R$_3$ and R$_4$ show a monolayer graphene-like dispersion. Despite this, R$_1$ and R$_2$ show a greater intensity than R$_3$ and R$_4$, even though states belonging to the bottom layer would be expected to be attenuated. In addition to this, replicas which we would expect to be equivalent with respect to their scattering distance from their respective primary show differing intensity. This is seen most clearly for the replicas R$_1$, R$_2$, R$_7$, R$_9$, R$_{12}$ and R$_{14}$. Although they are all a single moiré reciprocal lattice vector away from $\upkappa_2$, R$_7$, R$_9$, R$_{12}$ and R$_{14}$ are all at least an order of magnitude weaker than R$_1$ and R$_2$. The good agreement between experiment and theory confirms the validity of the model used for simulating the photoemission intensity.

\section{EDC analysis of hybridisation gaps}

\begin{figure}[H]
    \centering
    \includegraphics[width=0.7\textwidth]{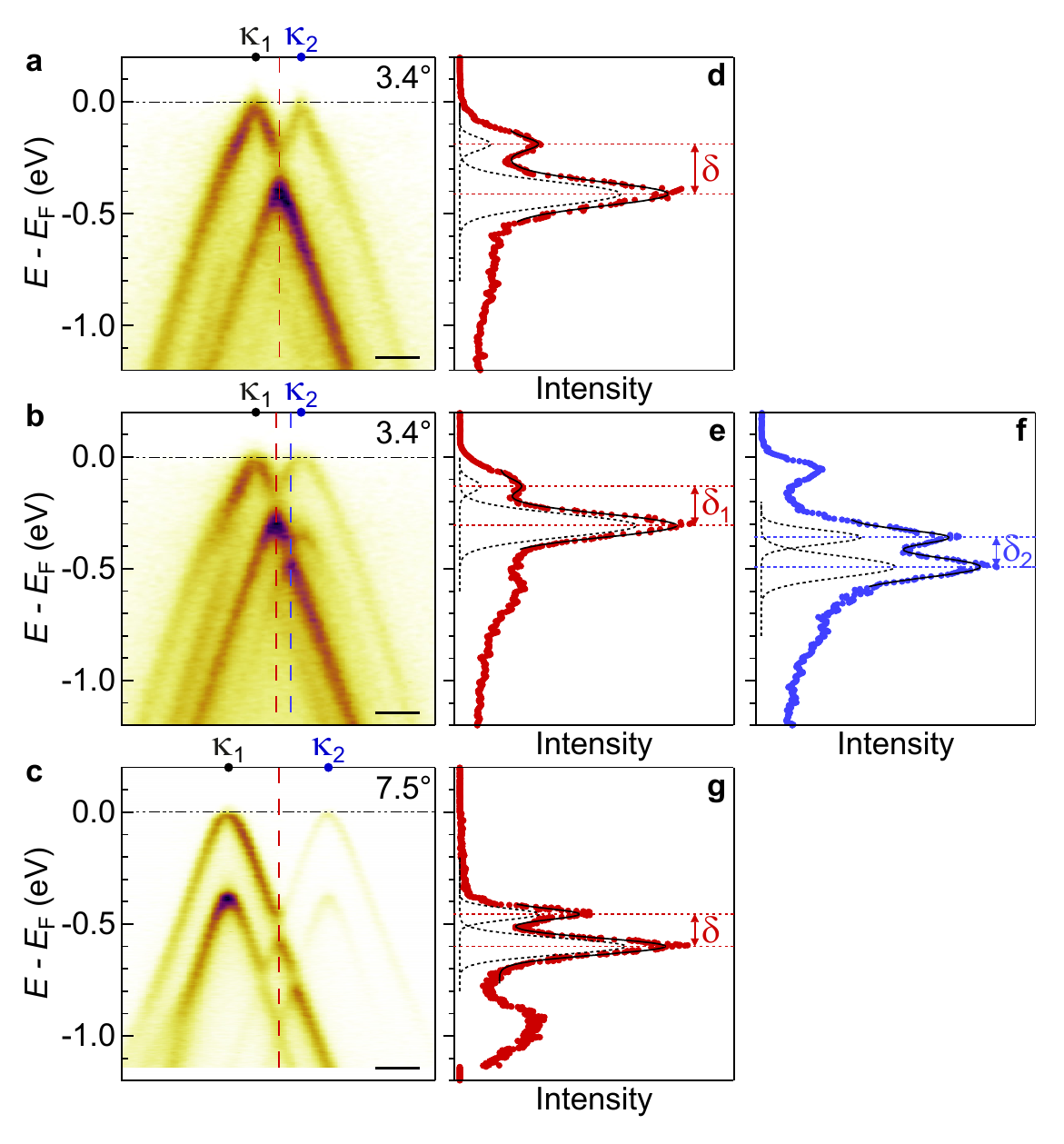}
    \caption{\textbf{EDC analysis of hybridisation gaps}. \textbf{a-c}~ARPES spectra along the $\upkappa_{1}$-$\upkappa_{2}$ direction for tBG, tMBG and tDBG, respectively, at the specified twist angles. \textbf{d-g}~EDCs extracted along the vertical dashed lines in \textbf{a-c} overlaid with a pair of Gaussians on a constant background fitting function. Dashed peaks show individual Gaussian fits, where the separation in their centres provides the hybridisation gap size, $\updelta$. All scale bars are 0.1~\AA$^{-1}$.}
    \label{Gaps}
\end{figure}

EDCs extracted from cuts along the $\upkappa_{1}$-$\upkappa_{2}$ direction were used to determine the size of hybridisation gaps. Examples are shown in Fig.~\ref{Gaps} for each of the different twisted graphene stacking arrangements discussed in the main text. Extracted EDCs are fit to a pair of Gaussian functions on a constant background (Fig.~\ref{Gaps}d-g). The difference between the Gaussian peak positions is interpreted as the hybridisation gap size, $\updelta$.

\section{tDBG flatband}

\begin{figure}[H]
    \centering
    \includegraphics[width=\textwidth]{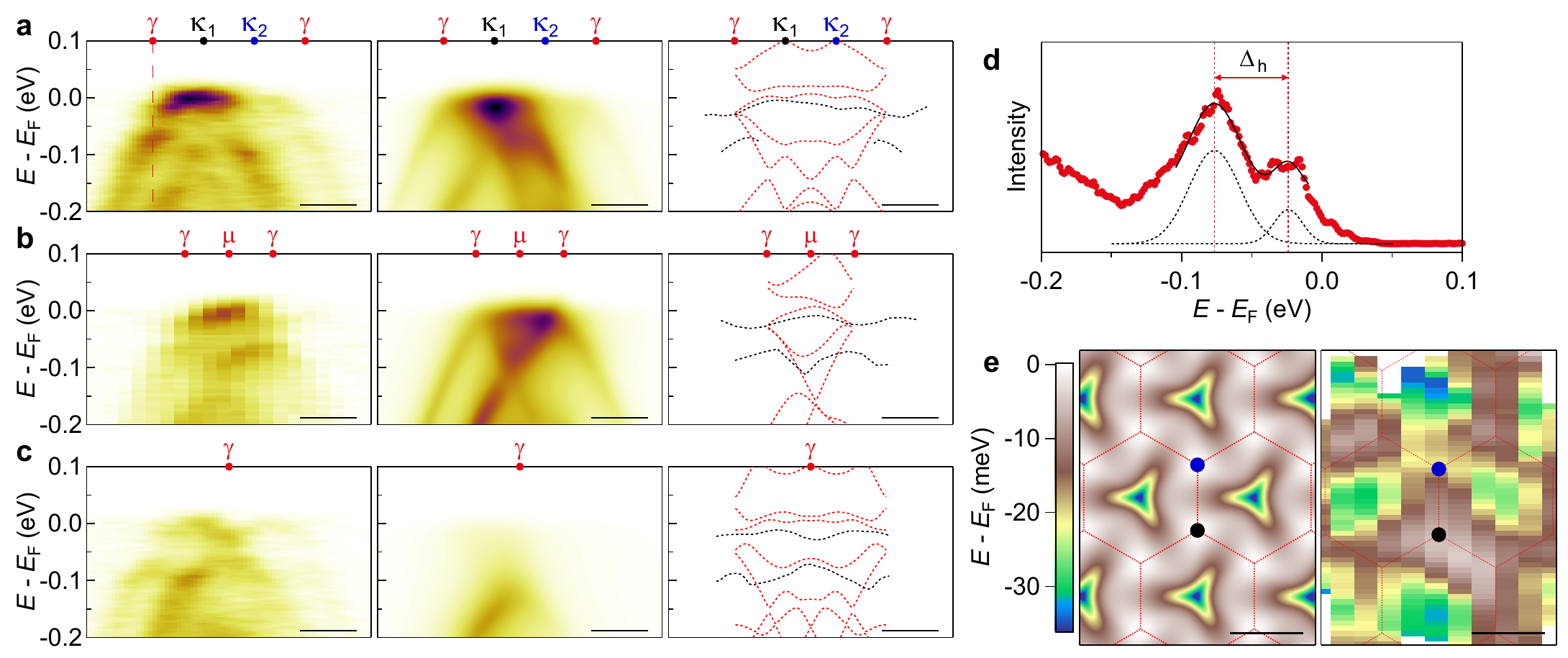}
    \caption{\textbf{Simulated and experimental ARPES spectra of 1.5$^\circ$ tDBG}. \textbf{a-c}~Experimental (left-hand) and simulated (middle) energy-momentum cuts along the high symmetry directions, as in Fig.~3 of the main text, and the corresponding band dispersions (right-hand). The black lines correspond to the peak positions extracted from the experimental data by fitting EDCs and the red lines correspond to the predicted electronic structure. \textbf{d}~EDC along the vertical dashed line in the left-hand panel of \textbf{a}. The solid black line is a fit to the data with a pair of Gaussian peaks whose positions correspond to the band positions and separation gives the gap size, $\Delta_h$. \textbf{e}~Energy of the simulated (left-hand) and experimental (right-hand) flat-band plotted in the $\textit{k}_\textit{x}-\textit{k}_\textit{y}$ plane, with the mBZs overlaid in red. All scale bars are 0.05~\AA$^{-1}$.}
    \label{Flatband}
\end{figure}

Fig.~\ref{Flatband} is an extended version of Fig.~3 from the main text including further comparison with the continuum model close to $E_\text{F}$. There are clear differences between the simulated ARPES spectra in the middle panels of Figs.~\ref{Flatband}a-c and the corresponding experimental spectra. Inspection of the band dispersions (right-hand panels) reveals that the gap between the upper valence band (the flat band) and the lower-lying valence bands is significantly smaller in the predicted dispersions (red lines) than the experimental band dispersions (black lines). Note that the 60~meV broadening applied to the simulated spectra in the main text is reduced to 40~meV here to match the improved quality of the experimental results attained for this device. Fig~\ref{Flatband}d illustrates how the experimental band dispersions were obtained: the EDC is from the red dashed line in the left-hand panel of Fig~\ref{Flatband}a, the solid black line is a fit using two Gaussian peaks the positions of which give the band dispersions. In Fig.~\ref{Flatband}e, the energy of the flatband is plotted in the $k_x-k_y$ plane across the first few mBZs for both the continuum model calculated dispersion (left-hand panel) and the dispersion extracted from the experimental spectra (right-hand panel). Though the features in the experimental data are broad, they agree with the prediction that the flat band minima should be at the $\gamma$ point and show the expected periodicity across the mBZs.

\section{Self-consistent analysis of the effect of a back-gate voltage}

The effect of a back gate voltage is included in the electronic structure model through a self-consistent analysis that accounts for the change in interlayer potential due to the displacement field and the resultant charge redistribution. The electric displacement field has the following form~\cite{Xu2021}:
\begin{equation}
    D=\frac{V_\text{G} C_\text{G}}{2 \epsilon_0},
\end{equation}
where $C_\text{G}$ is the capacitance to the back gate \cite{Guo2019}. This can be used to calculate an initial interlayer potential~\cite{Slizovskiy2021}, $u$, that is the difference between the potential on the top layer, $U_t$, and the bottom layer, $U_b$:
 \begin{equation}
     u_i = U_t-U_b = \frac{e c_0 D}{\epsilon_0 \epsilon_z},
     \label{Delta}
 \end{equation}
 where $c_0$ is the spacing between the layers and $\epsilon_z$ is the effective out-of-plane dielectric susceptibility. For tMBG, where there are three layers, the energy differences between the two outer layers and the inner layer must be calculated. For this, as in~\cite{Slizovskiy2021}, the following parameters are used:
 \begin{equation}
     \begin{split}
        & c_{0,1}=3.44~\text{\AA~for tBLG}, \\
        & c_{0,2}=3.35~\text{\AA~for BLG}, \\
        & \epsilon_{z,1}= 2.5~\text{for tBLG}, \\
        & \epsilon_{z,2} = 2.6~\text{for BLG}. \\
     \end{split}
 \end{equation}
 The initial interlayer potential $u_i$ is introduced to the Hamiltonian as follows:
 \begin{equation}
     H \longrightarrow H + 
     \begin{pmatrix}
      0 & 0 & 0 & 0 & 0 & 0 \\
      0 & 0 & 0 & 0 & 0 & 0 \\
      0 & 0 & -u_1 & 0 & 0 & 0 \\
      0 & 0 & 0 & -u_1 & 0 & 0 \\
      0 & 0 & 0 & 0 & -(u_1+u_2) & 0 \\
      0 & 0 & 0 & 0 & 0 & -(u_1+u_2) \\
     \end{pmatrix}.
 \end{equation}
 
From the wavefunctions calculated using this new Hamiltonian, the new layer density $n_i$ in each layer due to the back gate is calculated ($n_t$ for the density in the upper monolayer graphene, $n_m$ for the density in the middle layer, and $n_b$ for the density in the bottom layer of the bilayer graphene), as are the energies of each layer: 
 \begin{equation}
          n = \frac{\epsilon_0 \epsilon_\text{hBN} V_\text{G}}{d_\text{hBN}e} ,
          \label{N}
 \end{equation}
 \begin{equation}
              n_i = 2 \int_{BZ} \frac{d^2k}{(2\pi)^2} \sum_{l=1}^{2N}\left[ (  |\Psi^l_{A_i}  (k)|^2+ |\Psi^l_{B_i}(k)|^2)f(\epsilon_l-E_F)-\frac{1}{2} \right].
              \label{ni}
 \end{equation}
 Here, $\epsilon_\text{hBN}=4$ \cite{hbn}, and $d=26$~nm is the thickness of the hBN layer for the data in Fig.~5 of the main text. In Eq.~(\ref{ni}), the index $i$ denotes the layer and $N$ is the number of bands being considered for the calculation. $\Psi_{\lambda}$ are the wavefunctions and $f(\epsilon_l-E_\text{F})$ is the Fermi distribution. The wavefunctions are used to calculate a new set of interlayer energy differences and the calculations are iterated until they converge and the interlayer potentials are found self-consistently. For tMBG, these give the interlayer potential between the monolayer and the upper layer of the bilayer graphene, $u_{1}$, and between the upper and lower layers of the bilayer graphene, $u_{2}$:
\begin{equation}
    \begin{split}
        & u_{1}(D,n)=\frac{eD_z c_{0,1}}{\epsilon_0 \epsilon_{z,1}} + \left[\frac{e^2(n_t-n_m)}{2 \epsilon_0} \frac{1+\epsilon_{z,1}^{-1}}{2} - \frac{e^2n_b}{2 \epsilon_0 \epsilon_{z,1}}\right]c_{0,1} \\
        & u_{2}(D,n)=\frac{eD_z c_{0,2}}{\epsilon_0 \epsilon_{z,2}} + \left[\frac{e^2(n_m-n_b)}{2 \epsilon_0} \frac{1+\epsilon_{z,2}^{-1}}{2} + \frac{e^2n_t}{2 \epsilon_0 \epsilon_{z,2}}\right]c_{0,2}
    \end{split}
\end{equation}
 Eq.~(\ref{ni}) is then used to calculate the layer densities shown in Fig.~5c of the main text.

\section{Analysis of gated tMBG Dirac cones}

\begin{figure}[H]
    \centering
    \includegraphics[width=0.95\textwidth]{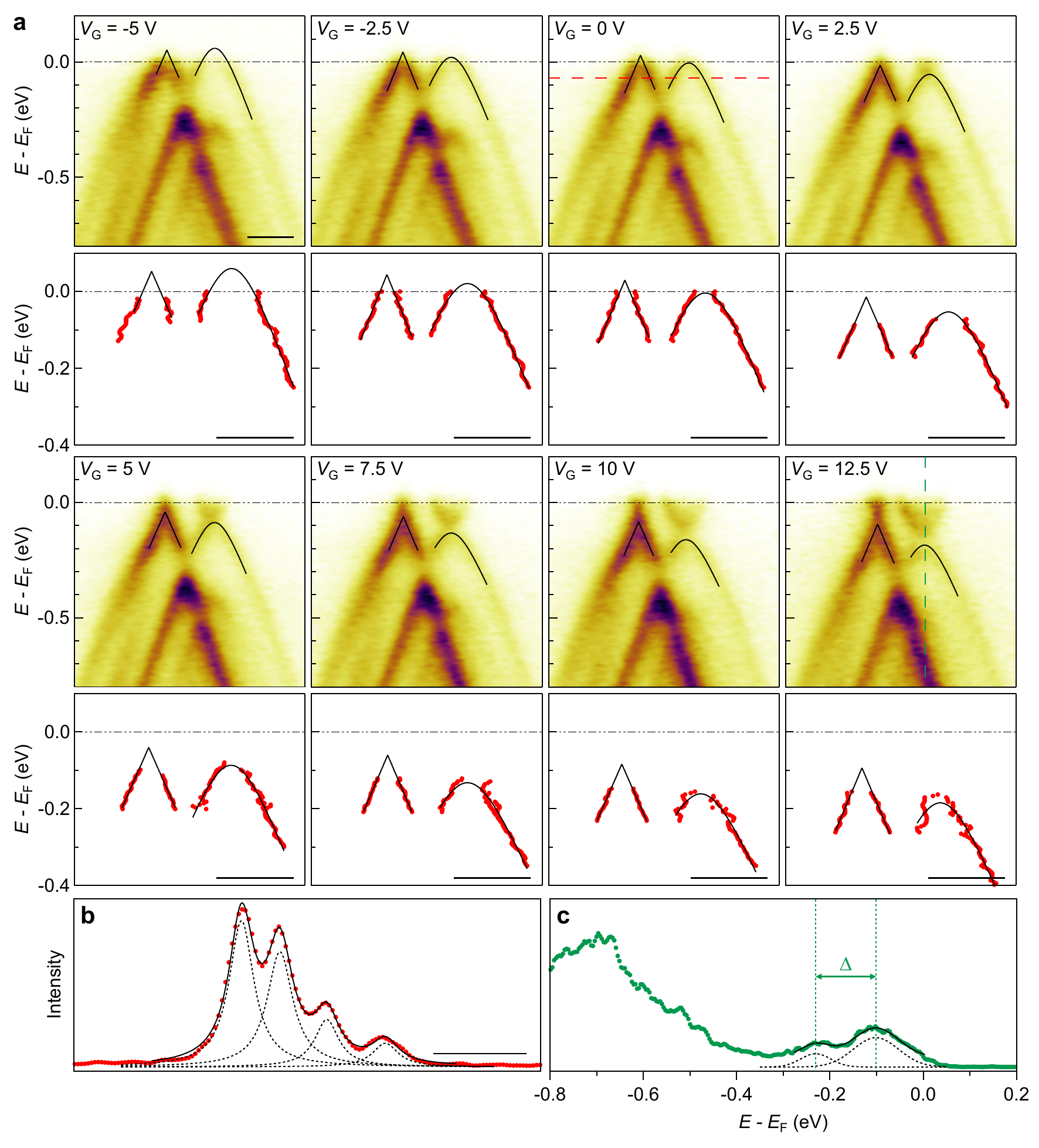}
    \caption{\textbf{Analysis of tMBG spectra at different gate voltages}. \textbf{a}~ARPES spectra along the $\upkappa_{1}$-$\upkappa_{2}$ direction for tMBG (top panel) and extracted band positions for the monolayer and bilayer cone (bottom panel) at different gate voltages. Solid lines are fits to the extracted band positions using the low-energy dispersion relations for monolayer and bilayer graphene. \textbf{b}~MDC extracted along the red horizontal dashed line in the $\textit{V}_{\text{G}} = 0~\text{V}$ spectrum. Dashed peaks show individual Lorentzian fits to each band. \textbf{c}~EDC extracted along the vertical dashed line in the $\textit{V}_{\text{G}} = 12.5~\text{V}$ spectrum. Dashed peaks show individual Gaussian fits, the separation between their peak energies provides the bilayer gap size, $\Delta$. All scale bars are 0.1~\AA$^{-1}$.}
    \label{Gating}
\end{figure}

To calculate the band parameters as a function of $V_\text{G}$ from the experimental spectra, as plotted in Fig.~5 of the main text, energy momentum cuts through the $\upkappa_{1}$-$\upkappa_{2}$ direction were extracted for each gate voltage (Fig.~\ref{Gating}a top panels). Momentum distribution curves (MDCs) were used to extract the band positions of the monolayer and bilayer cones close to the Dirac points. These were fit to Lorentzian functions on a constant background, with the peak centre providing the band position (Fig.~\ref{Gating}b). Using standard low energy approximations to the electronic dispersions, the monolayer band positions were fit by $E = E_{\text{D}}-v|k-k_0|$, and the bilayer band positions by $E = E_{\text{D}}-\frac{1}{2}\gamma_1\left[\sqrt{1+4v^2k^2/\gamma_1^2}-1\right]$, Fig.~\ref{Gating}a bottom panels, where $v$ is a band velocity. The fitting coefficients provide the Dirac point energy $E_{\text{D}}$. From the tight-binding approximation to the isolated layers, the Dirac point energy can be used to calculate the carrier density, using expressions $n_\text{MLG}=\frac{E_{\text{D}}^2}{\pi v^2}$ and $n_\text{BLG}=\frac{\gamma_1E_{\text{D}}}{\pi v^2}$ for the monolayer and bilayer, respectively,~\cite{McCann2011}. For simplicity, we have only fit to the valence band, and thus assume the monolayer and bilayer dispersions are symmetric about $E_{\text{D}}$.

The size of the gap at the Dirac point of the bilayer graphene, $\Delta$, is measured in the same way as previously described  for the hybridisation gaps, see section S7. An EDC is extracted through the centre of the bilayer cone and fit to a pair of Gaussian functions on a constant background (Fig.~\ref{Gating}c). The gap can only be resolved for $V_\text{G} \geq 7.5$~V.

\printbibliography